\documentclass[sigconf,authorversion,screen]{acmart}

\usepackage{booktabs}
\usepackage{algorithm}
\usepackage{algorithmic}
\usepackage{graphicx}
\AtBeginDocument{%
  \providecommand\BibTeX{{%
    \normalfont B\kern-0.5em{\scshape i\kern-0.25em b}\kern-0.8em\TeX}}}

\setcopyright{acmcopyright}
\copyrightyear{2025}
\acmYear{2025}
\acmDOI{XXXXXXX.XXXXXXX}

\acmConference[SIGMOD '25]{In 2025 International Conference on Management of Data}{June 22-27,
  2025}{Berlin, Germany}
\acmBooktitle{SIGMOD '25: the International Conference on Management of Data,
 June 22-27, 2025, Berlin, Germany} 
\acmPrice{15.00}
\acmISBN{978-1-4503-XXXX-X/18/06}

\usepackage{hyperref}

\usepackage{fancyhdr}
\usepackage{multirow} % for \multirow
\usepackage{array}
\newcolumntype{C}[1]{>{\centering\arraybackslash}p{#1}}

\usepackage{amssymb}

\usepackage{bbm}
\usepackage{amsmath,amsfonts,amsthm}
\newtheorem*{remark}{Remark}
\usepackage{textcomp}
\usepackage{subcaption}

\theoremstyle{definition}

\newenvironment{exmp}
  {\begin{example}}
  {\hfill$\square$\end{example}}
\usepackage[multiple]{footmisc}
\newcommand{\checkcomment}[1]{{\color{black}{#1}}}

\usepackage{makecell}
\definecolor{tabblue}{rgb}{0.12156862745098039, 0.4666666666666667, 0.7058823529411765}      % tab:blue
\definecolor{tabmyorange}{rgb}{1.0, 0.4980392156862745, 0.054901960784313725}                   % tab:orange
\definecolor{tabgreen}{rgb}{0.17254901960784313, 0.6274509803921569, 0.17254901960784313} % tab:green
\definecolor{tabpurple}{rgb}{0.5803921568627451, 0.403921568627451, 0.7411764705882353}         % tab:purple
\definecolor{tabred}{rgb}{0.8392156862745098, 0.15294117647058825, 0.1568627450980392}    % tab:red
\definecolor{tabbrown}{rgb}{0.5490196078431373, 0.33725490196078434, 0.29411764705882354}   % tab:brown
\begin{document}

% \pagestyle{plain}
% \title{\textsc{LITune}: An End-to-End Automatic Tuning System for Learned Index}
\title{A New Paradigm in Tuning Learned Indexes: A Reinforcement Learning Enhanced Approach}
% \title{Tuning Learned Indexes: A Reinforcement Learning Approach}
% %
% % The "author" command and its associated commands are used to define
% % the authors and their affiliations.
% % Of note is the shared affiliation of the first two authors, and the
% % "authornote" and "authornotemark" commands
% % used to denote shared contribution to the research.
% \affiliation{%

% \affiliation{%

% \affiliation{%

\author{Taiyi Wang}
\affiliation{%
  \institution{University of Cambridge}
  \city{Cambridge}
  \country{United Kingdom}
}
\email{Taiyi.Wang@cl.cam.ac.uk}

% \author{Liang Liang}{Imperial College London, epfl}{liang.liang20@imperial.ac.uk, liang.liang@epfl.ch}{https://orcid.org/0000-0002-4566-6178}{}

\author{Liang Liang}
\affiliation{%
  \institution{EPFL}
  \city{Lausanne}
  \country{Switzerland}
}
\email{liang.liang@epfl.ch}
\authornote{Work has been partly performed at Imperial College London, London, UK.}

% \email{liang.liang20@imperial.ac.uk,
% \\ liang.liang@epfl.ch}

\author{Guang Yang}
\affiliation{%
  \institution{Imperial College London}
  \city{London}
  \country{United Kingdom}
}
\email{guang.yang15@imperial.ac.uk}

\author{Thomas Heinis}
\affiliation{%
  \institution{Imperial College London}
  \city{London}
  \country{United Kingdom}
}
\email{t.heinis@imperial.ac.uk}

\author{Eiko Yoneki}
\affiliation{%
  \institution{University of Cambridge}
  \city{Cambridge}
  \country{United Kingdom}
}
\email{eiko.yoneki@cl.cam.ac.uk}
\authornote{Corresponding author.}

\begin{abstract}

%new version:
Learned Index Structures (LIS) have significantly advanced data management by leveraging machine learning models to optimize data indexing. However, designing these structures often involves critical trade-offs, making it challenging for both designers and end-users to find an optimal balance tailored to specific workloads and scenarios. While some indexes offer adjustable parameters that demand intensive manual tuning, others rely on fixed configurations based on heuristic auto-tuners or expert knowledge, which may not consistently deliver optimal performance.

This paper introduces \textsc{LITune}, a novel framework for end-to-end automatic tuning of Learned Index Structures. \textsc{LITune} employs an adaptive training pipeline equipped with a tailor-made Deep Reinforcement Learning (DRL) approach to ensure stable and efficient tuning. To accommodate long-term dynamics arising from online tuning, we further enhance \textsc{LITune} with an on-the-fly updating mechanism termed the O2 system. These innovations allow \textsc{LITune} to effectively capture state transitions in online tuning scenarios and dynamically adjust to changing data distributions and workloads, marking a significant improvement over other tuning methods.  Our experimental results demonstrate that \textsc{LITune} achieves up to a 98\% reduction in runtime and a 17-fold increase in throughput compared to default parameter settings given a selected Learned Index instance. These findings highlight \textsc{LITune}'s effectiveness and its potential to facilitate broader adoption of LIS in real-world applications.

\end{abstract}

\maketitle
% \vspace{-5pt}
\section{Introduction}
\label{sec:intro}

The intersection of data management and machine learning has given rise to learned index structures. These indexes integrate machine learning, replacing traditional algorithmic components, to capture data distributions and optimize search times. Notable examples include  RMI \cite{kraska2018case}, ALEX \cite{ding2020alex} and PGM \cite{ferragina2020pgm}, etc., which have become subjects of extensive research.

The effective design of a learned index involves deliberate trade-offs to achieve optimal performance for varying workloads. For instance, ALEX favors combined search and update performance by introducing gaps at the expense of space efficiency \cite{ding2020alex}. On the other hand, the dynamic PGM Index prioritizes update efficiency over search performance \cite{ferragina2020pgm}. These design trade-offs also lead to more complex structures which generate configurable parameters. Tuning these parameters is the key to balancing the trade-offs that ensure higher performance over traditional indexes.

Beyond the primary parameters, learned indexes like ALEX have more subtle tunable factors that are often overlooked for simplicity. These parameters affect various aspects of the index performance, from operation cost (e.g., search and insertion cost) to the structure of the index (e.g., heights of the tree). For example, for ALEX, the \textit{Max Node Size} parameter changes the size of the nodes, thereby affecting the height of the tree. On the other hand, \textit{Split Policy} and \textit{Gap Ratio} affect how insertion is carried out. These parameters are intertwined, and adjusting them in real-world scenarios can lead to substantial performance improvements, though it requires a more complex tuning process.

\begin{figure*}[ht]
\centering
    \includegraphics[width=1.0\linewidth]{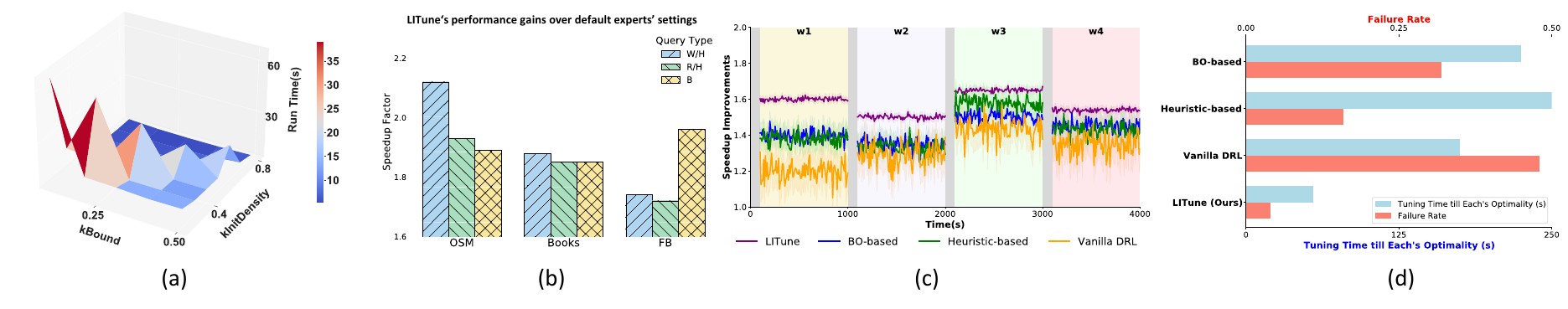}
    % Try adjusting the trim values {left bottom right top} until the empty space is minimized
    \vspace{-22pt} % Adjust space between the figure and the caption
\caption{(a) shows the performance surface of a learned index (ALEX) under a wild exploration of the parameter space. (b) highlights the optimal performance speedup achieved by \textsc{LITune} compared to default expert-selected parameters. (c) illustrates the continuous tuning performance of our system alongside other out-of-the-box methods under default configurations. (d) compares the tuning stability and costs across methods to reach their respective optimal performance levels.}
\label{Perf}
\vspace{-5pt} % Adjust space below the caption
\end{figure*}

\begin{figure}[ht]
\captionsetup{labelfont={bf,color=black}}
\centering
    \includegraphics[width=1.0\linewidth]{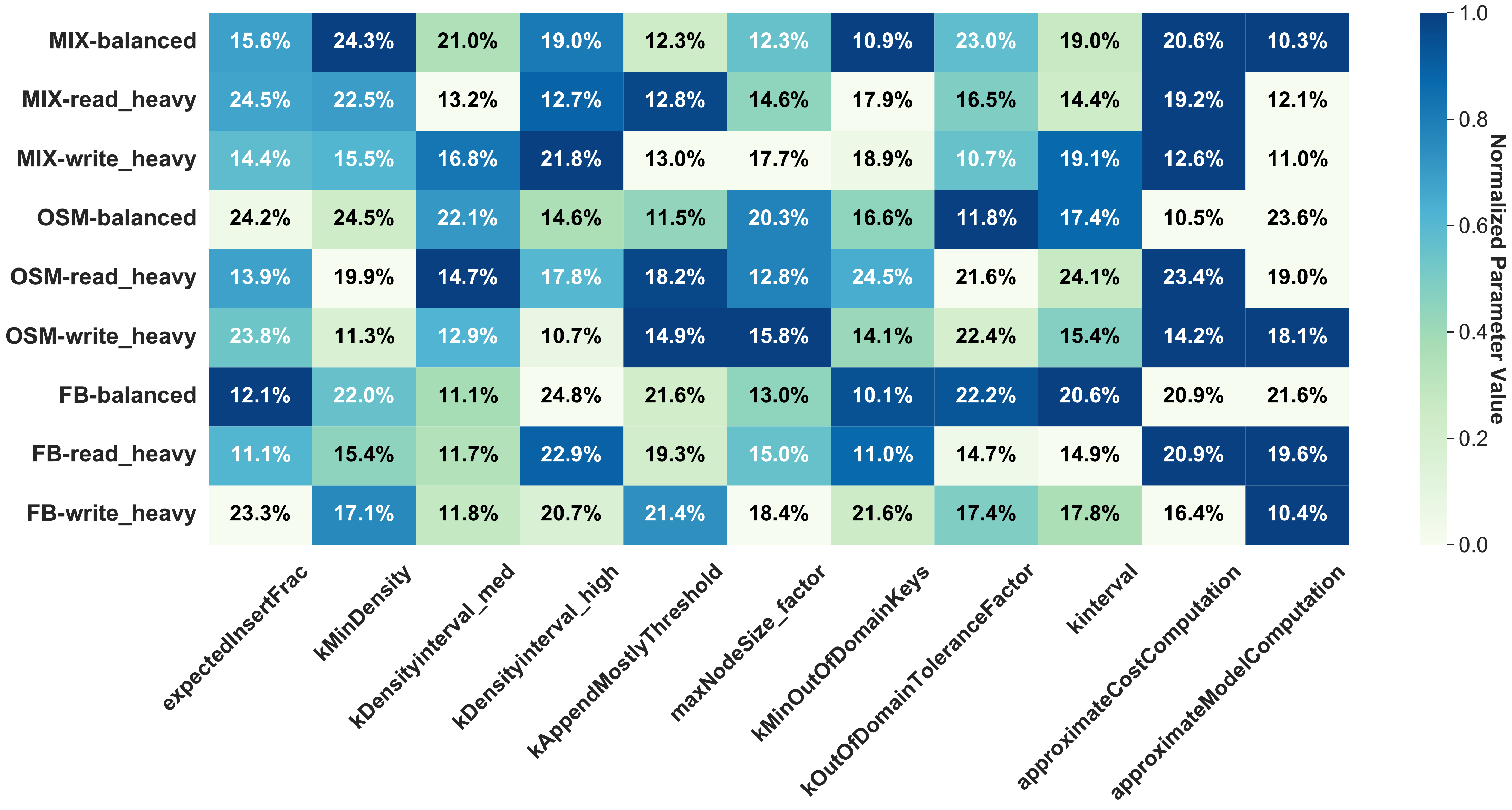}
    % Try adjusting the trim values {left bottom right top} until the empty space is minimized
    \vspace{-22pt} % Adjust space between the figure and the caption
\caption{Selected parameter value distributions and their impact scores across different workloads when tuning on ALEX. The heatmap colors represent normalized optimal parameter values, while the percentages indicate each parameter's individual tuning impact. }
\label{fig:p_impact}
% \vspace{-15pt} % Adjust space below the caption
\end{figure}

Selecting the right tuning approach for learned indexes involves navigating a myriad of parameter configurations~\cite{khalil2017learning, wang2020neural}. For example, in practice, parameterized indexes can exhibit vastly different performance due to parameter choices. This is illustrated in Figure~\ref{Perf}(a), where adjusting just two parameters leads to significant variability in runtime\footnote{In our experiments shown in Figure~\ref{Perf}(a) and (d), we executed 16 million write and 16 million read queries on a 1 million SOSD dataset~\cite{kipf2019sosd} using ALEX~\cite{ding2020alex}, selectively varying two parameters to validate the substantial performance gap among parameters.}. This variability underscores the complexities and potential performance swings when considering the full spectrum of high-dimensional and continuous parameter configurations, which can scale into thousands or millions. This complexity is compounded by the fact that a misconfiguration can lead to drastic performance deviations, emphasizing the importance of precise tuning. \checkcomment{Besides, our empirical experiments demonstrate that parameter interactions in learned indexes exhibit complex, workload-dependent relationships with no dominant parameters. As shown in Figure~\ref{fig:p_impact}, where colors represent normalized parameter values and percentages show impact scores (ratio of individual-parameter to full-parameter tuning improvements), no parameter consistently exerts greater influence, with all impact scores falling between 10-25\%. This heterogeneous distribution of parameter values across workloads, coupled with the balanced impact scores, indicates that performance optimization requires holistic parameter tuning rather than focusing on individual parameters.}  Moreover, unlike algorithmic indexes such as B+trees that perform well out of the box, learned indexes are distribution-dependent. Furthermore, systems are not designed to automatically tune indexes and typically do not allocate substantial resources for this purpose. However, there is a critical requirement for indexes to perform optimally; this \hypertarget{c1}{\textbf{necessitates that the tuner identifies high-quality solutions within a limited budget} (\textbf{Challenge C.1})}.

% this \hypertarget{r2-o3-3}{}\textbf{necessitates that the tuner identifies high-quality solutions within a limited budget} (\textbf{Challenge C.1})\label{challenge_c1}.   

% By fine-tuning these parameters, we can achieve significant performance improvements. Subfigure~(b) in Figure~\ref{Perf} emphasizes the substantial performance gains (measured by query runtime speedup) achieved by our tuning system over SOSD~\cite{kipf2019sosd}, via comparisons between the optimal solution found by \textsc{LITune} and default settings from experts. This demonstrates that \textbf{tuning is necessary} and can lead to considerable enhancements.

% While learned indexes strive to be user-friendly by abstracting away underlying parameters, this abstraction can inadvertently lead to significant performance degradation, as hiding these critical parameters from the user may result in suboptimal default settings for various scenarios. By fine-tuning these parameters, we can achieve significant performance improvements. Figure~\ref{Perf}.(b) emphasizes the substantial performance gains (measured by query runtime speedup) achieved by our tuning system over SOSD~\cite{kipf2019sosd} through comparisons between the optimal solution found by \textsc{LITune} and the default settings provided by experts. This demonstrates that \textbf{\textit{tuning is necessary}} and can lead to considerable enhancements. 

While learned indexes strive to be user-friendly by abstracting away underlying parameters, this abstraction can inadvertently lead to significant performance degradation, as hiding these critical parameters from the user may result in suboptimal default settings for various scenarios. By fine-tuning these parameters, we can achieve significant performance improvements. Figure~\ref{Perf}(b) emphasizes the substantial performance gains (measured by query runtime speedup) achieved by our tuning system over SOSD~\cite{kipf2019sosd} through comparisons between the optimal solution found by \textsc{LITune} and the default settings provided by experts. \checkcomment{Importantly, unlike many DBMSs whose default configurations are often overly conservative~\cite{zhang2019end,van2017automatic}, the default settings in learned indexes are decided based on instance-optimized designs~\cite{ding2022sagedb} which allow them to adapt to specific system environments given a "good" tuning algorithm. Currently, these algorithms are chosen by experts to optimize for common system environments. However, there hasn't been any end-to-end tuning system that is adaptive across data distributions.} Thus, \textbf{\textit{tuning is necessary}} and can lead to considerable enhancements.

% Our approach with \textsc{LITune} addresses this issue by automatically identifying and adjusting these parameters, ensuring both usability and high performance without requiring manual intervention from the user.

Moreover, in real-world usage, data distributions and query types are not constant, which makes tuning more challenging. Figure~\ref{Perf}(c) shows the degradation of tuning performance among existing out-of-box tuning methods during continuous online tuning when facing dynamic workloads, highlighting \textbf{\hypertarget{c2}the need for adaptive tuning to workloads and data distributions} (\textbf{Challenge C.2}). We introduce four different workloads derived from mixture-distributed data from the SOSD dataset~\cite{kipf2019sosd}, involving various query types by adjusting the read-write ratio over time. To ensure fair comparisons, each tuning method is provided with a preparation period, depicted as grey areas within the workload intervals. During this period, a preliminary smaller dataset reflecting future trends is used for warming up, initial tuning will happen during this period to make sure the system begins with reasonably optimized parameters.

Another challenge arises from the need for safe tuning, especially when dealing with a large parameter space and concurrent tuning demands. \checkcomment{Recently, Reinforcement Learning enhanced by deep neural networks (DRL) has already been proved by many works~\cite{kamthe2018data,zhang2019end} as a good tuner when working within a large parameter space due to its intrinsic exploration abilities. Equipped with a learned module, an RL-based approach can be easily generalized and deployed to various data conditions. However, DRL, as a trial-and-error-based approach, normally presents aggressive tuning towards the optimal solutions, whose potential risks to the existing system were largely ignored.} In such cases, \textbf{\hypertarget{c3}it is crucial to ensure tuning remains both safe and stable during exploration within the extensive parameter space} (\textbf{Challenge C.3}). Figure~\ref{Perf}(d) shows the failure rates caused by improper parameter settings from aggressive tuning approaches, particularly those resulting from exploratory vanilla DRL methods (PPO~\cite{schulman2017proximal}). We also present the differences in tuning costs until each method reaches optimality, further emphasizing the motivation for introducing \textsc{LITune}. These observations underscore the necessity of a tailor-made tuning system rather than relying on out-of-box methods.

The \textsc{LITune} system, utilizing a tailored Deep Reinforcement Learning (DRL) framework, significantly improves the tuning of learned index parameters, overcoming the constraints of traditional methods and optimizing performance without requiring the extensive training data needed by supervised models. Key contributions include:

(1) \textbf{Introduction of an Automatic Tuning System Using DRL (Addressing \fcolorbox{blue}{white}{\hyperlink{c1}{C.1}})}: This system efficiently navigates large parameter spaces in real-time, dynamically adjusting to changing performance requirements, capturing the stateful transitions, thus rapidly finding optimal solutions in complex online environments.

(2) \textbf{Adaptive Design (Addressing \fcolorbox{blue}{white}{\hyperlink{c2}{C.2}})}: During the training stage, we leverage Meta-RL to efficiently transfer knowledge from a smaller pre-training set to unseen tuning tasks through solid initialization and fast gradient descent. In the online tuning stage, \textsc{LITune} employs an on-the-fly updating system, termed the O2 system, to further boost adaptability over the longer term. This design enhances the tuner's ability to quickly adapt to new or evolving workloads and data distributions.

(3) \textbf{Operational Stability and Reliability (Addressing \fcolorbox{blue}{white}{\hyperlink{c3}{C.3}})}:  Featuring a Context-RL-based risk mitigation strategy (ET-MDP solver), \textsc{LITune} avoids dangerous configurations, ensuring the tuning process maintains the system's operational stability and reliability.

To our best knowledge, \textit{\textbf{\textsc{LITune} is the first system to enable stateful, online tuning of learned indexes using tailored Deep Reinforcement Learning methods, integrated with safety-aware mechanisms and O2 system to ensure reliable and adaptive performance.}}

The paper is organized as follows: Section~\ref{sec:related_works} reviews related work. Section~\ref{sec:sys} discusses our motivation and the design of \textsc{LITune}. Section~\ref{sec:method} details our novel RL-based tuning methodologies. Section~\ref{sec:Exp} presents our experimental analysis and results. Finally, Section~\ref{sec:con} concludes with a summary and discussion.

\section{Related Works}
\label{sec:related_works}
% This section reviews various works on parameter and index tuning. We emphasize the techniques in automatic parameter tuning specific to this domain.

% \begin{table*}[ht]
%     \centering
%     \begin{tabular}{l l p{9cm}}
%         \hline
%         \textbf{Method} & \textbf{Parameter Selection/Tuning Method} & \textbf{Weakness and differences} \\
%         \hline
%         B-trees \cite{comer1979ubiquitous} & Fixed & No search for optimal fanout \\
%         RMI \cite{kraska2018case} & Fixed & No hyperparameter tuning, fixed two-layer structure \\
%         ALEX/APEX \cite{ding2020alex} & Grid-search and expert selection & No end-to-end latency optimization \\
%         CDFShop \cite{marcus2020cdfshop} & Heuristic-based searches; Pareto front & Inconclusive tuning, not latency-optimized \\
%         AirIndex \cite{chockchowwat2023airindex} & Heuristic + graph-based & Optimizes lookup over updates, high tuning costs and limited scalability  \\
%         RusKey \cite{mo2023learning} & Vanilla DRL (DDPG) & Tuned specifically for LSM-trees, not generalizable, high tuning costs \\
%         Ours & Safe RL approach & Stable and efficient tuning under shifts of workloads and data \\
%         \hline
%     \end{tabular}
%     \caption{Summary of Existing Indexing and Tuning Works}
%     \label{tab:index_methods}
%     \vspace{-15pt}
% \end{table*}

\subsection{Parameter Tuning for System}
\label{sec:PT}
Parameter tuning is a common practice to optimize systems. For example, knob tuning in database systems, where specific values or knobs can be adjusted to optimize for specific query operators and improve data access efficiency~\cite{zhao2023automatic}. Traditional search strategies include using random and grid search to find the optimal parameters for a given workload. Advanced frameworks such as GPTune \cite{liu2021gptune}, Spearmint \cite{snoek2012practical}, and Sequential Model-Based Optimization (SMBO) \cite{ozaki2020multiobjective} attempt to refine this process through Bayesian Optimization, integrating multi-task and transfer learning. However, these methods struggle with accuracy and computational efficiency, as they require starting anew with each shift in workload or data distribution. The inefficiencies amplify when faced with real-time tuning requirements for learned indexes~\cite{feurer2015initializing}.

\subsection{Deep Reinforcement Learning for Parameter Tuning}
\label{sec:TD}
Recently, Deep reinforcement learning (DRL) has shown promising results in optimizing complex systems under dynamic conditions~\cite{kamthe2018data,chua2018deep}. Notably, CDBTune \cite{zhang2019end} uses deep deterministic policy gradients (DDPG~\cite{lillicrap2015continuous}) to automatically navigate and tune the high-dimensional continuous space of database configurations. It has shown the adaptability and efficiency of DRL methods over traditional tuning tools and expert DBA interventions for handling dynamic conditions. However, CDBTune cannot easily adapt to the unique challenges in index tuning, not present database configuration tuning. Index tuning requires rapid and precise adjustments without system resets or prolonged downtime while answering queries and updating records within dynamic data distributions and workload patterns. Misconfiguration has severe consequences for performance and must be avoided. \textsc{LITune}, on the other hand, efficiently tackles these challenges while keeping the full advantages of DRL tuning.

% \subsection{Index Tuning}
% Tuning indexes with traditional methods has been studied for years~\cite{chockchowwat2023airindex}. Recently, reinforcement learning techniques have also been incorporated into tuning index structures. RusKey \cite{mo2023learning} employed reinforcement learning to optimize an LSM-tree-based key-value store. Gu et al.\cite{gueffectively} applied reinforcement learning techniques to optimize R-Tree. These existing works demonstrate the successful application of RL to index tuning; our work, \textsc{LITune}, seeks to extend this to learned index tuning.

% \vspace{-5pt}

% \vspace{-5pt}
\subsection{Learned Index Tuning}

Learned indexes \cite{kraska2018case} replace traditional indexing algorithms (like B+Trees) with models that predict the approximate location of a key using the Empirical Cumulative Distribution Function (CDF), potentially reducing search operations. Research on both static \cite{kraska2018case, kipf2020radixspline, ferragina2020pgm, stoian2021towards} and updatable indexes \cite{galakatos2019fiting, ferragina2020pgm, ding2020alex, zhang2021carmi, wu2021updatable, yang2023flirt, liang2024swix, li2021finedex, tang2020xindex} demonstrates that performance heavily depends on data distribution, and to achieve the best performance, the indexes must be "tuned" according to the data distribution.

Currently, there are two approaches for considering data distributions when designing learned indexes: (1) exposing parameters for adjustment by users or future research, as seen in \cite{kraska2018case, ferragina2020pgm, zhang2021carmi}, and (2) implementing self-tuning mechanisms through cost models, utilized by indexes such as \cite{ding2020alex, yang2023flirt, liang2024swix}. In both approaches, the default parameter settings are crucial, with index designers asserting that tuning is unnecessary for satisfactory performance. However, this assumption only holds when the empirical cost of the default parameters is effective \cite{lu2021apex, li2021finedex, tang2020xindex}, and for updatable learned indexes, the insert cost model must also be valid \cite{ding2020alex, liang2024swix}. Early studies like \cite{sun2023learned} use grid search to find optimal default parameters but fail to capture the complexities of tuning learned indexes. Automating this tuning process remains under-explored and is fundamental to our work with \textsc{LITune}. This challenge is exacerbated for indexes supporting dynamic workloads, which must be reorganized to maintain model accuracy as data distributions shift. Unlike traditional indexes, learned index parameters depend not only on dataset size but also on rapidly changing data distributions. Additionally, updatable indexes require structural modifications—such as gaps \cite{ding2020alex, liang2024swix}, hierarchical structures \cite{ferragina2020pgm, wu2021lipp}, and buffers \cite{galakatos2019fiting, liang2024swix}—to mitigate the impact of distribution shifts. These complexities create intricate dependencies between parameters (as shown in Figure \ref{Perf}(a)), making automated tuning particularly challenging.

{\color{black}
CDFShop~\cite{marcus2020cdfshop} automates learned index optimization by fine-tuning cumulative distribution function (CDF) models specifically for the RMI index, adjusting high-level hyperparameters (e.g., model type, branching factors). However, it is confined to RMI, lacks safety-aware mechanisms, and relies on iterative, game-theoretic, and heuristic exploration methods that can be time-consuming and less scalable for complex indexes or larger datasets. Consequently, we exclude CDFShop from our baselines, comparing \textsc{LITune} instead with more general Bayesian Optimization and heuristic-based methods. AirIndex~\cite{chockchowwat2023airindex} employs a graph-based approach to automatically tune learned index structures but often has limited exploration by focusing on top-$K$ heuristics, incurring high overhead despite parallelization and complicating optimization in high-dimensional parameter spaces. Recently, Reinforcement Learning (RL) has been applied to tuning learned indexes: RusKey~\cite{mo2023learning} optimizes LSM-tree-based key-value stores~\cite{sutton2018reinforcement}, marking the first RL-driven LSM-tree transformations under dynamic workloads. Unlike white-box cost models centered on I/O complexities, RusKey’s black-box RL approach offers a holistic view similar to \textsc{LITune}~\cite{dayan2018dostoevsky}, yet it tunes only a single parameter and thus cannot be directly used for general learned index tuning. 
}

To clarify our selection criteria and highlight the key tuning methods utilized in prior research, we have summarized relevant studies in Table~\ref{tab:index_methods}. We observe that irrespective of the specific index type being targeted, the underlying tuning methods are generally transferable across different contexts. Consequently, we have extracted these transferable methods and adopted them as baselines for our experiments. This approach allows us to systematically evaluate the effectiveness of our proposed tuning system against established methodologies.

\begin{table}[t]
    \centering
    \captionsetup{font=small}
    \small
    \begin{tabular}{l l p{9cm}}
        \hline
        \textbf{Method} & \textbf{Parameter Selection/Tuning Method}  \\
        \hline
        B-trees \cite{comer1979ubiquitous} & Expert selection,  Heuristics \\
        RMI \cite{kraska2018case} &  Expert selection,  Heuristics \\
        ALEX \cite{ding2020alex} & Heuristic cost model, Grid search \\
        SWIX \cite{liang2024swix} & Heuristic cost model, Grid search \\
        CARMI \cite{zhang2021carmi} & Expert selection, Hardware-aware \\
        CDFShop \cite{marcus2020cdfshop} & Heuristics; Pareto front \\
        AirIndex \cite{chockchowwat2023airindex} & Heuristics, Graph-based search  \\
        RusKey \cite{mo2023learning} & Vanilla DRL (DDPG)  \\
        Ours & Safe RL approach\\
        \hline
    \end{tabular}
    \caption{Summary of Existing Indexing and Tuning Works}
    \label{tab:index_methods}
    % \vspace{-18pt}
\end{table}

% \vspace{-5pt}
\subsection{Index Advisor}
Index advisors are tools used in physical database design to help administrators select optimal indexes at the schema level based on query workloads~\cite{chaudhuri1997efficient, valentin2000db2}. They focus on recommending which columns or combinations of columns should be indexed to improve query execution times and overall system efficiency. Recent RL-based index selection methods~\cite{kossmann2022swirl, zhou2024breaking, siddiqui2024ml, wang2024leveraging} continue this approach by automating the index selection process but still operate at the schema level. In contrast, our work operates on a fundamentally different level by tuning the internal parameters of learned index structures themselves rather than selecting which indexes to create. Learned indexes leverage machine learning models to predict data positions within a dataset~\cite{kraska2018case}, and their performance heavily depends on hyperparameters and model configurations. Traditional index advisors do not address the challenges associated with optimizing these internal parameters. Therefore, our DRL-based tuning framework enhances the performance of learned indexes through dynamic internal optimization, offering adaptability and efficiency improvements that traditional index advising methods do not provide. This distinction underscores that we are addressing a different problem space, focusing on the internal mechanics of learned indexes rather than on schema-level index selection.

\section{\textsc{LITune} System}
\label{sec:sys}

\subsection{Motivation}

Unlike existing index tuning works that concentrate on a limited set of observable parameters directly reflected in data structures, our work tackles the more complex challenge of tuning within a vast parameter space where parameters interact in non-independent and intertwined ways. This complexity demands stable and efficient tuning strategies, particularly in the context of online and continuous learning tasks. Since parameter metrics are difficult to capture and do not readily lend themselves to the integration of strong heuristics or expert knowledge, we focus on capturing stateful transitions and propose tailored end-to-end tuners. 

Furthermore, navigating complex parameter spaces poses significant challenges for traditional search strategies and advanced model-based approaches \cite{zhang2019end}. Traditional search strategies (random search, grid search, heuristics) fall short in navigating the extensive parameter space of learned indexes, while advanced out-of-box tuning frameworks (e.g., SMBO \cite{ozaki2020multiobjective}) require starting a new navigation cycle with each shift in workload or data distribution, making them resource-intensive. Furthermore, establishing universal states for different index structures with different parameter sets creates complications. In this regard, \textsc{LITune} is designed to offer online and stateful index tuning using deep reinforcement learning for dynamic workloads and provides fast and safe configurations for multiple learned indexes. Our approach mitigates the instabilities and aimless explorations that often accompany the use of generic, out-of-the-box tuning methods.

\subsection{System Overview}

% This section outlines our automated learned index tuning system, \textsc{LITune}, which utilizes Deep Reinforcement Learning (DRL) to enhance performance. 

At the core of \textsc{LITune} is reinforcement learning (RL), which enhances adaptability to dynamic workloads beyond the capabilities of traditional cost models. Specifically, we design a Markov Decision Process framework that involves the RL agent interacting with an environment to maximize rewards. The decisions are made based on observed \textit{states} and chosen \textit{parameters} \cite{sutton2018reinforcement}, which, in this case, are the structural responses to shifting data distributions and tuned parameters, respectively. 

\textsc{LITune} operates in two primary phases: the Training Stage and the Online Tuning Stage. In the Training Stage, we implement an efficient adaptive training pipeline to generate a generalizable pre-trained model. Once this model is deployed, it undergoes continuous fine-tuning in the Online Tuning Stage, ensuring it remains current and effective under evolving operational conditions. To avoid misconfigurations during tuning, \textsc{LITune} adopts a context-aware RL system that prevents early terminations, ensuring stability and speed throughout the process. Additionally, to keep the RL agent updated during online use, we employ an O2 system for ongoing training and adjustments.
% The detailed methodology of incorporating Context-Aware RL into learned index tuning will be further elaborated in Section~\ref{sec

% In the context of learned index tuning, ensuring operational safety is a paramount challenge. To robustly address this, we frame the tuning process as an Early-Terminated Markov Decision Process (ET-MDP)~\cite{sun2022constrained}, wherein tuning actions that potentially violate safety constraints lead to immediate termination of the process. This strategy draws on the principles of Constrained Markov Decision Processes (CMDPs)~\cite{wang2019benchmarking}, embedding safety directly into the decision-making framework.

\begin{figure*}[!t]
\centering
    \includegraphics[width=\linewidth]{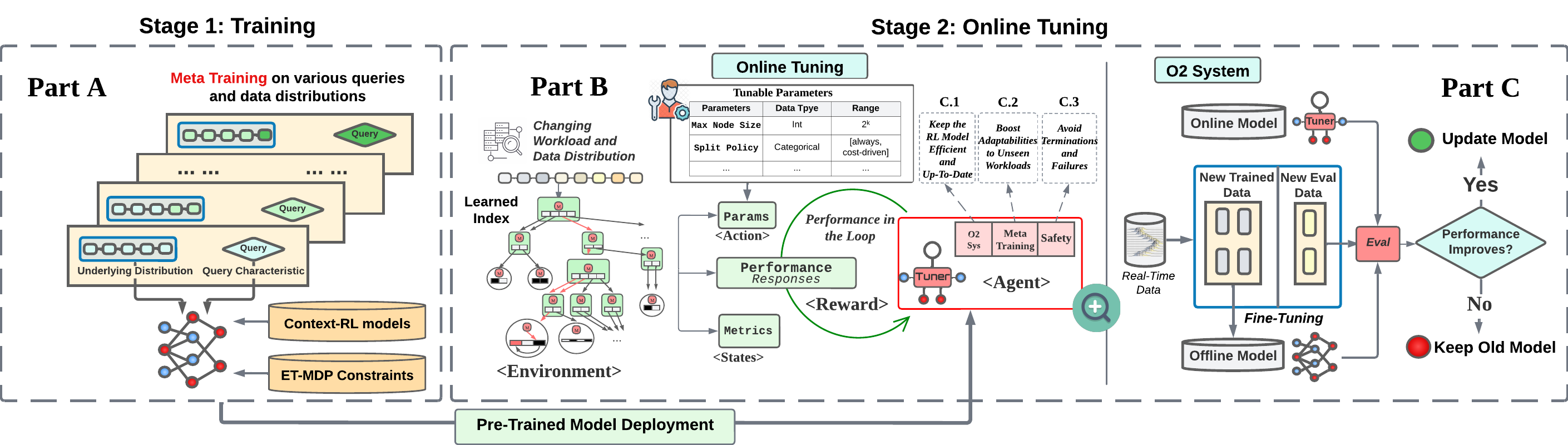}
\caption{The architecture of \textsc{LITune}. \textbf{Part A} illustrates the training phase, where RL-based models are trained. Once the training is complete, these models are deployed as online tuners in \textbf{Part B}. The operational details of the O2 system are explained in \textbf{Part C.}}
% \vspace{-10pt}
\label{LITune}
\end{figure*}

\subsection{Training Stage}
\label{sec:pre-training}

Part A of Figure \ref{LITune} depicts the initial generation of a pre-trained RL agent. This agent is crucial as it forms the foundation of our adaptive RL-based tuner, designed to efficiently handle diverse tuning scenarios right from deployment.

\vspace{-3pt}
\subsubsection{Offline Training Preparation}

The foundational step in our approach involves preparing a Deep Reinforcement Learning (DRL) agent for optimizing learned index configurations. This process commences with the generation of varied datasets and query sets, establishing a diverse training environment. Each training episode, defined by a unique dataset-query set combination, allows the agent to explore a spectrum of configurations through a trial-and-error strategy, thereby accumulating a rich set of initial training data.

\textbf{Training Data:} Unlike supervised learning, our methodology relies on the automatic collection of training quintuples \( \langle d, q, a, s, r \rangle \) by the RL agent. \( d \) denotes the dataset, \( q \) a set of queries, \( a \) the parameters for index construction, \( s \) the index states, and \( r \) the performance metric. This approach ensures comprehensive feedback for the agent, which is pivotal for its learning process. Given the NP-hard nature of optimizing learned index configurations in a continuous parameter space, our model employs general DRL techniques. This choice enables the exploration of novel configurations and mitigates the risk of entrapment in local optima.

\subsubsection{Adaptive Training (Meta Training) Design}
\label{sec:adaptive-training}

As mentioned in Introduction,  the \textbf{\fcolorbox{blue}{white}{\hyperlink{c2}{C.2}}} requires robust generalization of  Deep Reinforcement Learning (DRL) agents in \textsc{LITune} to handle real-world scenarios with unseen queries and variable data distributions~\cite{ying2018transfer}. To achieve this, we introduce an adaptive training design based on Meta Reinforcement Learning (Meta-RL), which enhances the tuner's adaptability to new situations and addresses. 

% We emphasize the training process in detail because many automatic tuners built on learned models omit these crucial steps, making them difficult to reproduce. 

Our method employs the Model-Agnostic Meta-Learning (MAML) approach~\cite{fu2023maml2}, which trains agents to rapidly adapt with minimal updates. In the context of learned index tuning, the "tuning instances" represent specific scenarios with unique data distributions and query types. MAML integrates into the RL training process through a two-level training loop:

\textit{Inner Loop—Adaptation to Tuning Instances:} Agents perform tuning-instance-specific updates to optimize performance on sampled tuning instances. This involves adjusting policy parameters based on interactions characterized by specific workload types and data distributions.

\textit{Outer Loop—Meta-Update for Generalization:} The initial policy parameters are updated across all tuning instances to improve general performance. This enhances the agent's ability to generalize to new, unseen tuning scenarios.

This dual-loop process enables the agent to handle individual tuning scenarios effectively while maintaining broad adaptability across diverse operational conditions, which is essential for efficiency in the dynamic landscape of learned index tuning. Here is a quick example on how it works in practice:

% \footnote{Due to the space limit, the pseudo-code for implementation to our Meta-RL application is outlined in our supplementary materials here: \url{https://github.com/SubAnony/LITune_SIGMOD_25/tree/master/supplemental_materials}} 

\begin{exmp}
    \textbf{Practice of Meta-RL}
    
    Consider two tuning instances for a learned index system:
    
    \begin{enumerate}
        \item \textit{Instance A}: Primarily range queries on a uniformly distributed dataset.
        \item \textit{Instance B}: A mix of insert and range queries on a skewed, non-uniform dataset.
    \end{enumerate}
    
    A traditional RL agent trained only on Instance A performs well for similar uniform range queries. However, when applied to Instance B, it struggles to maintain performance, requiring extensive retraining to adjust its policy. In contrast, a Meta-RL agent trained with the MAML approach has encountered diverse tuning instances during meta-training. When faced with Instance B, it can swiftly adapt its policy with just a few gradient updates, leveraging prior knowledge. This enables the Meta-RL agent to maintain robust performance across varying scenarios without significant retraining. 
    % \square
\end{exmp}

\vspace{-3pt}
\subsection{Online Tuning Stage}
\label{sec:online}

After establishing a strong foundation in the training stage, we now transition to the online tuning stage, where the pre-trained model is put into practical use. This section describes how the \textsc{LITune} system operates and adapts to continuous tuning needs on the fly.

Parts B and C of Figure \ref{LITune} illustrate the architecture of the online tuning system. The dashed box at the top represents the client and data storage system where end users send their queries to the \textsc{LITune} system below. This system is designed to handle the continuous adaptation required by varying data environments.

The operational flow is presented in Part B of Figure~\ref{LITune}: Once the data storage setup and tuning requests are confirmed, the learned index and tuning system process the current data as the underlying distribution. The system executes queries on this data, generating performance data and states. Based on these observations, the tuning system automatically suggests adjustments to the learned index parameters to optimize future query handling and improve performance.

Furthermore, \textsc{LITune} leverages the pre-trained model to provide real-time recommendations for parameter settings during online tuning scenarios. It also continuously refines this model by incorporating feedback from each tuning request into its training process. This integration of Online Tuning and Offline Training (O2 System) ensures that \textsc{LITune} dynamically adapts to any changes in workload and data distribution, maintaining high efficiency and adaptability over time.

% \vspace{-8pt}
\subsubsection{Online Tuning}
End users can easily tune the target learned index by submitting a request to \textsc{LITune}. Upon receiving a request, the system gathers the necessary data and utilizes the pre-trained RL model for online tuning, ultimately recommending the best-performing parameters.

% \vspace{-8pt}
\subsubsection{O2 System}
\label{sec:O2}

As depicted in Part C of Figure~\ref{LITune}, the O2 system integrates Online and Offline RL models to address \textbf{\fcolorbox{blue}{white}{\hyperlink{c2}{C.2}}}, enhancing real-time adaptability and performance optimization. The system employs the online tuner with a pre-trained model for immediate index adjustments when no data changes occur. Conversely, significant data changes activate both the online and offline models: the offline model refines itself with new data, while the online model handles real-time optimizations.

The O2 system routinely assesses the necessity for model updates by comparing the online model's performance against new data and predefined criteria, including statistical divergence and user-defined thresholds. This assessment occurs at regular intervals or user-defined checkpoints, ensuring the O2 system remains responsive to changing data and workloads. During updates, the system carefully balances the current online model's performance with the insights gained from the offline model's continuous learning, thus maintaining optimal tuning efficacy. This efficiency also contributes to addressing \textbf{\fcolorbox{blue}{white}{\hyperlink{c1}{C.1}}}.

It should be noted that the real-world queries encountered during online tuning might differ from those used in the offline training process. For this reason, this structure allows \textsc{LITune} to stay adaptive, leveraging the offline system's incremental fine-tuning for better alignment with real-world workloads, while the online system provides swift responses to immediate tuning needs. Here is a practical working example of O2 system:

\begin{exmp}
    \textbf{Running O2 System}
    
    Consider a learned index tuning scenario with two distinct phases of data workload:
    
    \begin{enumerate}
        \item \textit{Stable Phase}: The dataset experiences minimal changes, and the workload consists mainly of read-heavy range queries.
        \item \textit{Dynamic Phase}: The dataset undergoes significant updates, including frequent insertions and deletions, and the workload shifts to a mix of read and write queries.
    \end{enumerate}
    
    During the \textit{Stable Phase}, the O2 system utilizes the online tuner with the pre-trained model to make immediate index adjustments, ensuring efficient query processing without the overhead of model retraining.
    
    When transitioning to the \textit{Dynamic Phase}, the significant data changes trigger the activation of both Online and Offline models. The offline model begins refining the tuning strategy by learning from the new data distribution and varied query patterns. Simultaneously, the online model continues to handle real-time optimizations to maintain query performance.
\end{exmp}

\subsection{\textsc{LITune} Working Process} 

Summarizing the learned index parameter tuning in \textsc{LITune}, the learned index (tuning target) serves as the RL environment, while the deep RL model acts as the agent, recommending configurations based on the learned index state. As the index is constructed and queries are executed, the learned index state alters, reflected in the metrics. These metrics evaluate learned index performance, calculating the corresponding RL reward value, and the agent updates its policy accordingly, continuing until the tuning time budget is exhausted, ultimately revealing the most fitting parameter settings. Thus, the RL-based tuner can easily capture and memorize the state responses to the data distribution and workload shifts.

To illustrate how our RL-based tuner is specifically designed for learned index scenarios, Figure~\ref{exp+Safe}(a) demonstrates a tuning example using ALEX. As the workload transitions from a balanced to a write-heavy workload, \textsc{LITune} detects changes in ALEX's states and performance metrics. Initially, it increases the maximum node size from the default 16MB and reduces the threshold for out-of-domain inserts before triggering node expansion. This adjustment leads to a notable decrease in the "no\_expand\_and\_retrain" metric. With positive feedback from the training environment, \textsc{LITune} stores new data and asynchronously fine-tunes the model using the O2 system, evaluating potential updates based on pre-defined criteria. This enables continuous learning with minimal disruption, allowing swift adaptation to workload changes. Eventually, \textsc{LITune} recommends further increasing the maximum node size to 64MB and raising the minimum number of out-of-domain inserts required before expansion for future trials. The Safe RL module ensures system safety by automatically preventing the selection of dangerous states when configuring more aggressive values for the maximum node size and minimum number of out-of-domain inserts. We avoid these aggressive settings, which may yield immediate rewards but could lead to system failure in the long run. \checkcomment{While some learned indexes (like ALEX) require full reconstruction for structural parameter changes due to their codebase constraints, \textsc{LITune} provides flexible on-the-fly reconfiguration mechanisms across different index implementations. For cases where reconstruction is unavoidable, \textsc{LITune} minimizes overhead through efficient sampling: maintaining a small reservoir ($\approx 1\%$ of dataset) and proportionally selecting queries across workload types (read/write), only applying final configurations on the full dataset. This sampling strategy enables rapid performance estimation while preserving workload characteristics, with detailed cost analysis provided in Section~\ref{sec:tuning_training}.}

% \vspace{-5pt}
\section{Methodology}
\label{sec:method}

In this section, we explore the integration of novel Reinforcement Learning (RL) models to address the unique challenges of tuning learned indexes. While vanilla RL frameworks like Deep Deterministic Policy Gradient (DDPG) methods \cite{zhang2019end,lillicrap2015continuous} are adept at managing high-dimensional spaces and continuous actions, they often fall short in ensuring safety for tuning systems. To address these limitations, we have augmented the standard DDPG framework by incorporating context-aware learning. Notably, our online tuner components are designed with flexibility in mind and can be replaced by any existing vanilla DRL methods, showcasing our framework's adaptability to accept similar DRL approaches. Detailed discussions on these DDPG framework enhancements are presented in the subsequent sections.

\subsection{RL formalization for \textsc{LITune}}
\label{sec:rlform}
Using RL in \textsc{LITune} requires a nuanced formalization of learned index-tuning scenarios. Part B in Figure \ref{LITune} illustrates the interaction diagram of \textsc{LITune} components and the functional workflow of \textsc{LITune} in practice.

\begin{figure*}[t!]
\centering
    \includegraphics[width=0.85\linewidth]{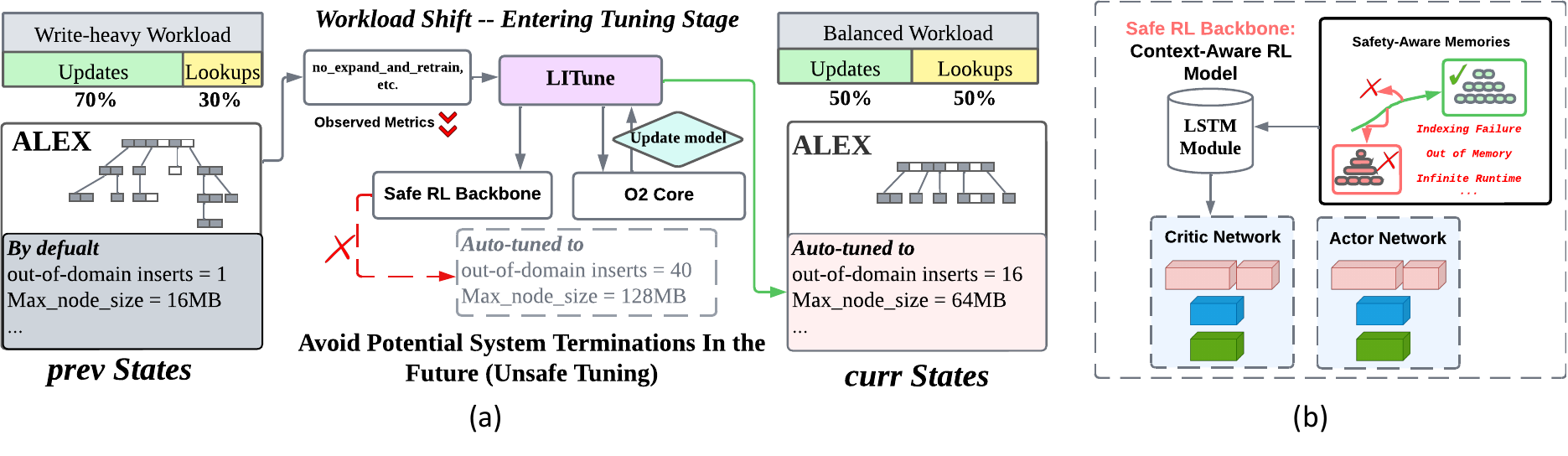}
    \vspace{-10pt}
\caption{(a) Running example of \textsc{LITune}. This example demonstrates how \textsc{LITune}'s tuner components respond to changes in performance metrics and adjust to workload shifts. (b) The safe-RL approach prevents aggressive tuning by learning from instabilities encountered during training.}
\vspace{-10pt}
\label{exp+Safe}
\end{figure*}

\textbf{Agent}: Viewed as the tuning system, the agent receives rewards and states from the learned index, updating the policy to steer parameter adjustments towards higher rewards.

\textbf{Environment}: Representing the tuning target, the environment is an execution instance of the learned index.

\textbf{State}: In \textsc{LITune}, the state of the agent, denoted as \( s_t \), represents the current condition of the learned index after applying recommended parameter settings. We categorize states into two main categories: structural and operational. Structural metrics might include features such as the number of internal nodes or tree height that represent the structure and measurable mechanisms of the index. However, not all features can be captured with just structural features. Therefore, we define a new category of operational metrics that captures which parameters affect the actual operation of the index. For each of the key operations (search, insert, update, and delete), we define a set of features that capture the cost of these operations using non-index-specific metrics. For example, we use search distance to capture the search cost, which is universal to all indexes regardless of the exact search algorithm used (linear, binary, or exponential). These metrics act as empirical proxies due to the complex inter-dependencies among features, offering a detailed snapshot of the index’s internal states necessary for effective tuning across various scenarios.

\checkcomment{
\textbf{Tuning-oriented Reward Design.} We define the reward $r_t$ as a scalar that accounts for performance changes from both the initial baseline ($D_0$) and the immediately preceding step ($t-1$). The idea is to capture the sort of incremental, forward-looking decisions that human experts make when tuning complex systems, balancing short-term gains with a broader trajectory of improvement. Concretely, the RL agent starts from an initial performance $D_0$ and seeks an improved state $D_n$. Once tuning begins, the system transitions to $D_1$ and the RL-agent computes $\Delta(D_1,D_0)$. From there, each iteration aims to surpass its predecessor, reflecting the principle that $D_i$ should exceed $D_{i-1}$ for all $i < n$. This design effectively encodes both immediate and foundational improvements in a single reward function.

Our focus metric for performance, denoted as \(R\), signifies the end-to-end runtime, which is paramount for understanding and optimizing query performance. To track optimization progress, we define two key differential metrics:
$\Delta =
\begin{cases}
\Delta_{t\to0} = \frac{-R_t + R_0}{R_0} \\
\Delta_{t\to t-1} = \frac{-R_t + R_{t-1}}{R_{t-1}}
\end{cases}$
Inspired by \cite{zhang2019end}, the reward, \(r\), is articulated as:
$$
\label{eq:rl}
    r = 
\begin{cases}
 ((1 + \Delta_{t\to0})^2 - 1)^\omega (1 + \Delta_{t\to t-1})^\kappa, & \text{if } \Delta_{t\to0} > 0 \\
-((1 - \Delta_{t\to0})^2 - 1)^\omega (1 - \Delta_{t\to t-1})^\kappa, & \text{if } \Delta_{t\to0} \leq 0
\end{cases}
$$
Here, \(\Delta_{t\to0}\) and \(\Delta_{t\to t-1}\) measure performance changes relative to the initial setting and the previous step. The scalar parameters \(\omega\) (odd) and \(\kappa\) (even) control how strongly the reward emphasizes near-term versus longer-term improvements: \(\omega\) governs the significance of changes from the initial baseline, and \(\kappa\) dictates the impact of recent performance gains or losses. By adjusting these scalars, practitioners can configure how aggressively the system pursues performance gains, how quickly it penalizes regressions, and how it balances near-term improvements against long-term objectives. In our practice, \(\omega=1\) and \(\kappa=2\) often strike a useful balance between achieving notable gains and maintaining stability.
}

\checkcomment{
\textbf{Multi-objective trade-offs and achieving the tuning goal via RL.}
The reward function underpins our RL-based tuning process by connecting the search mechanism to diverse performance objectives. By maximizing the expected cumulative discounted reward,
$
\max_{\pi} \mathbb{E}_{\tau \sim \pi} \Bigl[\sum_{t=0}^{T} \gamma^t r_t\Bigr],
$
where \(\pi\) is the policy, \(\tau\) the trajectory, \(\gamma\) the discount factor, and \(r_t\) the immediate reward, the RL agent explores parameter configurations aligned with user priorities. It refines its policy via trial-and-error (temporal-difference methods) to approximate expected returns across diverse states. Practitioners can adjust the performance metric \(R\) to emphasize or de-emphasize latency or throughput (e.g., \(R = 0.8 \cdot \text{latency} + 0.2 \cdot \text{throughput}^{-1}\)), steering the tuner’s optimization without altering the underlying RL framework, ensuring that the tuner can effectively meet the final goal—be it latency-sensitive, throughput-oriented, or a balanced mix of both.
}

\textbf{Action}: Denoted as \(a_t\), actions derive from the parameter configurations space and correspond to parameter tuning operations, influencing all tunable parameters concurrently.

\textbf{Policy}: Policy, mapping from state to action, maintains state transitions and is represented by a deep neural network. RL aims to learn the optimal policy.

\subsection{Backbone: Safe RL approach for \textsc{LITune}} 
\label{sec:LSTM-DDPG}

% As menetioned in introduction, we face the challenge of optimizing performance while ensuring safety (C3). Speciallacy we need to avoid configurations that lead to system instability or failures such as out-of-memory errors or endless runtime ....

As mentioned in the introduction, we face the challenge of optimizing performance while ensuring safety in the context of tuning learned index (\textbf{\fcolorbox{blue}{white}{\hyperlink{c3}{C.3}}}), i.e., avoiding configurations that lead to system instability or failures such as out-of-memory errors or endless runtime. To address this, we propose a minimalist approach by employing an \textit{Early Terminated Markov Decision Process} (ET-MDP) solver that triggers an \textit{early termination} whenever the learning policy violates predefined constraints. Early termination has been previously used to improve sample efficiency in solving regular MDPs~\cite{wang2019benchmarking}, as it accelerates learning by reducing the policy search space and shortening the time horizon. Moreover, an \textit{ideal} policy should \textit{never} violate the constraints, eliminating the need to learn to proceed or recover after violations.

To effectively handle the constraints in our tuning problem, we model it as a \emph{Constrained Markov Decision Process} (CMDP) and transform it into its early terminated counterpart, the ET-MDP~\cite{sun2022constrained}. This transformation allows us to apply standard RL algorithms while ensuring safety through the early termination mechanism.

\begin{definition}[Constrained Markov Decision Process]
\label{def:cmdp}
A \emph{Constrained Markov Decision Process} (CMDP) is a deterministic MDP with a fixed horizon $H \in \mathbb{N}^+$, defined by the tuple $(\mathcal{S}, \mathcal{A}, H, r, c, C, \mathbb{T})$, where $\mathcal{S}$ and $\mathcal{A}$ represent the state and action spaces; $r: \mathcal{S} \times \mathcal{A} \to \mathbb{R}$ is the reward function; $c: \mathcal{S} \times \mathcal{A} \to \mathbb{R}$ is the cost function representing constraints; $C \in \mathbb{R}^+$ is the upper bound on the permitted expected cumulative cost; and $\mathbb{T}: \mathcal{S} \times \mathcal{A} \to \mathcal{S}$ is the transition function. The policy class $\Pi$ consists of stationary policies $\pi: \mathcal{S} \times \mathcal{A} \to [0,1]$ such that $\sum_a \pi(a|s) = 1$ for all $s \in \mathcal{S}$.
\end{definition}

In our learned index tuning problem, constraints such as out-of-memory errors and endless runtime define dangerous or constrained states, as illustrated in Figure~\ref{exp+Safe}(b). We incorporate these constraints into the CMDP framework by defining appropriate cost functions. Specifically, we assign costs (e.g., $c_m$ for memory violations and $c_r$ for runtime violations) which are set to 1 upon violation, ensuring that each type of violation contributes equally to the cumulative cost. This approach penalizes the policy for entering unsafe areas and guides it toward safer trajectories.

\begin{remark}
\label{remark:cmdp-in-tuning}
By incorporating system constraints into the cost function, we can effectively model the safe learned index tuning problem as a CMDP.
\end{remark}

To handle the constraints and ensure safety during the learning process, we transform the CMDP into an \emph{Early Terminated MDP} (ET-MDP), which introduces an absorbing termination state whenever the cumulative cost exceeds a predefined threshold.

\begin{definition}[Early Terminated MDP]
\label{def:etmdp}
For any CMDP, we define its \emph{Early Terminated MDP} (ET-MDP) as a new unconstrained MDP $(\mathcal{S} \cup \{s_e\}, \mathcal{A}, H, r', \mathbb{T}')$, where $s_e$ is the absorbing state after termination. The transition function and reward function in the ET-MDP are adjusted to handle terminations:
\[
\mathbb{T}'(s, a) = \mathbb{T}(s, a)\mathbbm{1}(b_t \leq C) + s_e\mathbbm{1}(b_t > C),
\]
\[
r'(s, a) = r(s, a)\mathbbm{1}(b_t \leq C) + r_e\mathbbm{1}(b_t > C).
\]

where $b_t = \sum_{\tau = 1}^{t} c^m_{\tau} + c^r_{\tau}$ records the cumulative costs up to time $t$, and $r_e \in \mathbb{R}$ is a small termination reward. The parameter $C$ represents the total tolerated failures we can accept during training.
\end{definition}

These adjustments integrate the costs into the ET-MDP framework to ensure that once a constraint is violated, the associated cost is counted and added to the cumulative costs. This guides the agent to avoid actions leading to high-penalty states, effectively steering the policy towards safer and more optimal trajectories.

\begin{remark}
\label{remark:etmdp-in-tuning}
By converting the CMDP into an ET-MDP and solving it using an appropriate RL algorithm, we can effectively ensure that the learned policy respects the constraints by avoiding actions that lead to early termination.
\end{remark}

\textbf{\textit{Solving the ET-MDP:}}  The transformation of the CMDP into an ET-MDP simplifies the problem by converting it into an unconstrained MDP where constraints are implicitly handled via early termination. The goal is to find an optimal policy $\pi$ that maximizes the expected cumulative reward while respecting the constraints:
\[
\max_{\pi \in \Pi} \mathbb{E}_{\tau \sim \pi, \mathbb{T}} \left[\sum_{t=1}^H r_t\right], \quad \text{s.t.} \quad \mathbb{E}_{\tau \sim \pi, \mathbb{T}} \left[\sum_{t=1}^H c_t\right] \leq C,
\]
where $\tau = (s_1, a_1, r_1, \dots, s_H, a_H, r_H)$ represents the trajectory generated by policy $\pi$.

To solve this constrained optimization problem, the Lagrangian method relaxes it to an unconstrained one with a penalty term:
\begin{equation}
\label{eq:lagrangian}
\pi^{*}=\arg\max_{\pi\in\Pi}\min_{\lambda\ge0} \mathbb{E}_{\tau\sim\pi,\mathbb{T}}\left[\sum_{t=1}^H  r_t - \lambda \sum_{t=1}^H c_t\right] + \lambda C,
\end{equation}
Where $\lambda\ge0$ is the Lagrangian multiplier. In practice, if the policy $\pi$ is parameterized by $\theta$, i.e., $\pi = \pi_\theta$, the optimization over $\theta$ and $\lambda$ can be conducted iteratively through policy gradient ascent for $\theta$ and stochastic gradient descent for $\lambda$ according to Eqn.~(\ref{eq:lagrangian}).

However, as pointed out by \cite{chow2018lyapunov}, one possible defect of the Lagrangian methods is the violation of constraints during training, as the method may not strictly enforce the constraints at every iteration. This issue can be mitigated by incorporating \emph{context models}~\cite{sun2022constrained}.

{\color{black} Context models in an ET-MDP solver learn generalizable representations across similar tasks. In our setting, each state corresponds to a different task within the same distribution, allowing context models to transfer policies to unseen states and avoid constraint violations. By modeling tuning as a CMDP and transforming it into an ET-MDP, we naturally incorporate safe RL constraints into learned index tuning.}

\textit{\textbf{Implementation in \textsc{LITune}:}}
In \textsc{LITune}, we handle constraints such as memory limits and runtime bounds by triggering early termination when these constraints are violated. The normal CMDP solver, integrated with the Deep Deterministic Policy Gradient (DDPG) algorithm enhanced with Long Short-Term Memory (LSTM), allows the system to manage large state and action spaces effectively. The LSTM module maintains context from past explorations, enabling the RL agent to adapt to dynamic data environments and explore safely while maximizing reward collection.

Figure~\ref{exp+Safe} summarizes how the ET-MDP solver in \textsc{LITune} addresses safe tuning issues. The core strategy involves linking dangerous system \textit{\textbf{metrics}} to the learned index \textit{\textbf{states}} (understood as the structural information of the constructed index). The RL agent learns to avoid these dangerous states and maximize rewards within the given constraints during training. Once deployed, with memory units embedded in LSTM, the RL agent can autonomously identify safe tuning areas and achieve more reliable tuning performance. This strategy is significantly effective to handle the \textbf{\fcolorbox{blue}{white}{\hyperlink{c3}{C.3}}}.

% \vspace{-5pt}

\section{Experimental Study}
\label{sec:Exp}
%% Introduce the experiment designs

In this section, we evaluate the performance of \textsc{LITune} in comparison to existing tuning approaches, focusing on two specific learned index instances across various workloads and datasets. 

\subsection{Key Insights}

Before presenting our experimental results, we highlight several essential insights from our study:

\textbf{(a) Tuning is essential:} Effective default parameters are hard to set due to the complexity and variability of systems and problems. Unlike claims in~\cite{ding2020alex,zhang2021carmi}, default settings often fail to achieve optimal performance because they can't handle parameter inter-dependencies and dynamic data conditions. This highlights the crucial need for tuning to enhance learned index performance by synchronously adjusting parameters. (details in section \ref{sec:eff})

\textbf{(b) \textsc{LITune} is efficient:} Contrary to the belief that deep models sacrifice tuning efficiency for quality, \textsc{LITune} uses online tuning methods to significantly improve efficiency. Experiments demonstrate that \textsc{LITune} outperforms other methods in performance gains, regardless of the tuning budget. (details in section \ref{sec:E2E}, corresponding to \textbf{\fcolorbox{blue}{white}{\hyperlink{c1}{C.1}}})

\textbf{(c) \textsc{LITune} is adaptive:} \textsc{LITune} effectively handles online and continuous tuning scenarios, adapting to different data distributions and workload dynamics. Its ability to adjust to varying tuning budgets showcases its robustness and applicability across diverse index types, proving its effectiveness under changing data conditions. (details in section \ref{sec:adapt}, corresponding to \textbf{\fcolorbox{blue}{white}{\hyperlink{c2}{C.2}}})

\textbf{(d) \textsc{LITune} is safe:} While setting safe ranges for individual parameters is simple, adjusting multiple parameters simultaneously poses safety challenges. \textsc{LITune} employs a context-aware strategy that links failure situations with states, using constraints and penalties to maintain parameters within safe zones. This approach ensures system stability and safety during online tuning, as demonstrated by our experimental results. (details in section \ref{sec:safe_tune}, corresponding to \textbf{\fcolorbox{blue}{white}{\hyperlink{c3}{C.3}}})

% \vspace{-5pt}
\subsection{Experimental Settings}
\label{sec:exp_settings}

\subsubsection{Platform}

Experiments are conducted on a machine equipped with an NVIDIA Quadro RTX 8000 GPU, an Intel(R) Xeon(R) Gold 5218 CPU @ 2.30GHz, 8 vCPUs, and 64GB RAM. Since most of the parameter tuning methods have a certain degree of randomness, we repeated each experiment 5 times with different seeds.
\vspace{-5pt}
\subsubsection{Tuned Indexes and parameters}
\textbf{ALEX}~\cite{ding2020alex} is a dynamic, updatable learned index that allows tuning to optimize workload performance while not requiring parameters for basic operation. It offers statistics such as the number of model and data nodes, as well as expansions and splits, which serve as states for tuning parameters like read-write ratios, node sizes, and gap settings.

\textbf{CARMI}~\cite{zhang2021carmi}, based on RMI, enhances cache-awareness in learned indexes. It does not expose statistics directly but allows for state identification, including structural elements like leaf node counts and operational factors such as query visits and keys scanned. These states are tuned for parameters like leaf size and search operation weights, showcasing our method's capability to adapt tuning strategies to various learned index configurations.

We selected these two as our tuned instances because they exemplify two distinct parameterization mechanisms, each utilizing different optimization methods. Specifically, ALEX employs a heuristic-based cost model to automatically tune its parameters, whereas CARMI optimizes its parameters involving hardware-specific considerations.

\begin{table}[t]
\captionsetup{font=small,labelfont={bf,color=black}}
\caption{Parameter space characteristics of learned indexes}
% \vspace{-5pt}
\footnotesize
\label{tab:parameter_summary}
\centering
\small
\begin{tabular}{lcc}
\toprule
\textbf{Index} & \textbf{\# Dims} & \textbf{Parameter Types} \\
\midrule
ALEX & 14 & \makecell[l]{• 5 Continuous [0,1] (density, workload ratios,etc.) \\
                         • 3 Boolean (computation, splitting controls, etc.) \\
                         • 4 Integer (node sizes, buffer limits, etc.) \\
                         • 2 Discrete choice (tree policies, etc.)} \\
\midrule
CARMI & 13 & \makecell[l]{• 10 Continuous (operation timings, etc.) \\
                          • 2 Integer (node sizes, etc.) \\
                          • 1 Hybrid continuous/discrete (lambda, etc.)} \\
\bottomrule
\end{tabular}
\vspace{-10pt}
\end{table}

\checkcomment{Table~\ref{tab:parameter_summary} presents an overview of the parameter space in learned indexes, highlighting the marked differences in tuning complexity across existing methods. While other recent index tuning works typically handle only a few parameters---CDFShop~\cite{marcus2020cdfshop} focuses on 2--4 RMI-specific parameters, RusKey~\cite{mo2023learning} primarily tunes the compaction policy $K$ in FLSM-Tree (fewer than three parameters), and AirIndex~\cite{ding2020alex} employs layer-wise parameters which, despite their seemingly large number, are governed by strong hierarchical constraints, pruned to a top-$K$ set, predominantly focused on I/O-aware optimizations, and limited to static workload conditions---our system manages a substantially larger optimization space of around 10-15 parameters.
% \footnote{\label{fn:ref} \MySharedFootnoteText}

The complexity of our parameter space, with its high dimensionality and mixed discrete-continuous variables, poses significant challenges for traditional optimization methods under limited tuning budgets. RL-based methods excel by learning parameter interactions through experience, enabling efficient global exploration while avoiding local optima.}

\subsubsection{Evaluation and Training Datasets}

We evaluate \textsc{LITune} using the Search On Sorted Data (SOSD) benchmark suite~\cite{kipf2019sosd}, which includes datasets with up to 200 million 64-bit keys from various domains: Amazon books, OpenStreetMap (OSM), Facebook user IDs, and MIX (a combination of uniform, FB, books, and OSM distributions).

% \footref{fn:ref}

To prevent overfitting and ensure the tuner encounters unseen distributions during evaluation, we adopt strategies from prior RL-based training~\cite{zhang2019end,mo2023learning}. Instead of training directly on SOSD, we generate synthetic 1-D key-value pairs with diverse distributions (e.g., uniform, beta, normal) and vary the Write-Read Ratio (W/R Ratio) between 1:10 and 10:1. These experiments test the generalization ability of the methods across diverse query patterns in various tuning scenarios.

% enhance \textsc{LITune}’s ability to generalize across diverse query patterns and improve its adaptability and performance in varied tuning scenarios.

% \paragraph{Data Preparation and Training}

\subsubsection{Workloads}
\label{sec:workload}
\textsc{LITune} evaluation highlights its versatility across static and dynamic datasets, establishing it as a robust tuning solution for learned indexes. All runtime performances depicted in the section's figures represent the average runtime for basic operations (INSERT, SEARCH, and DELETE) based on the Write-Read Ratio within workloads.

{\color{black}
\textbf{(a) Static Workload:}
We evaluate \textsc{LITune} on static datasets (OSM, books, Facebook, MIX) to demonstrate its adaptability to varied data distributions. Our setup uses 90M records from a 100M dataset\footnote{A 90M subset aligns with the \textit{NIPS mlforsys'19} fb distribution in the SOSD benchmark.}, reserving 10M for a fixed distribution and 80M for INSERT/DELETE operations. Workloads differ by Write-Read Ratio (W/R): Balanced (1), Read-Heavy (1/3), and Write-Heavy (3). We evaluate tuning efficiency over varying steps (Figure~\ref{fig:tuning}) and also conduct \textbf{"extensive tuning"} (up to 50 steps or 1000 seconds) to gauge near-optimal performance (Figure~\ref{fig:Runtime_perf}).

\textbf{(b) Data-shifting Workload:}
We divide 100M records into 30 tumbling windows~\cite{patroumpas2006window}, each with 1M base data and 8M updates. \textsc{LITune} frequently adjusts parameters to handle rapid data evolution, with at most 5 tuning steps or 100 seconds (whichever first), as shown in Figures~\ref{fig:DS} and~\ref{fig:O2}\footnote{These workload combinations reflect their worst performance under streaming constraints.}.
}

\subsubsection{Evaluation Metrics}
\begin{figure*}[ht]
\centering
    \includegraphics[width=0.95\linewidth]{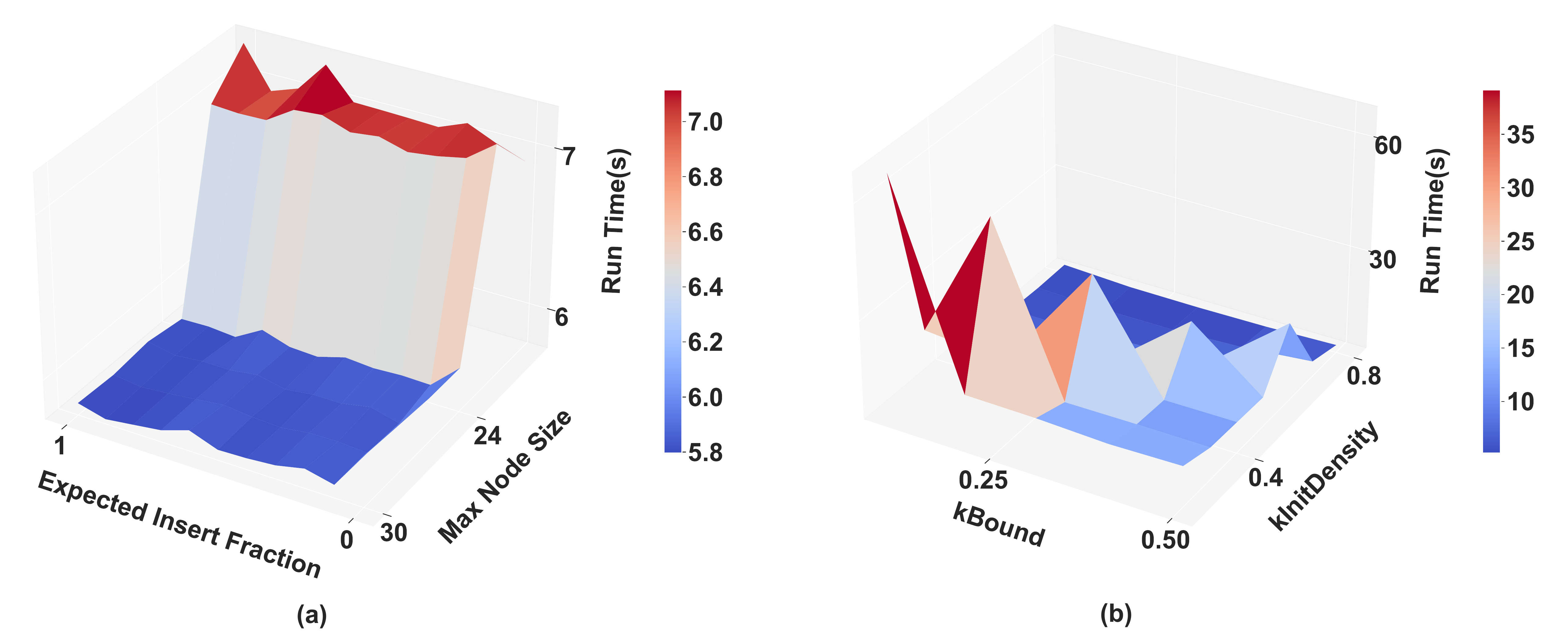}
\caption{Runtime performance (average on operation tuples) with extensive tuning. Performance improvements relative to default settings are marked.}
% \vspace{-5pt}
\label{fig:Runtime_perf}
\end{figure*}

The key metric used for evaluating the different tuning methods is End-to-end runtime performance. We crafted a dual-faceted experimental setup to encompass both static and data-shifting cases.  In addition to query runtime performance shown in figure~\ref{fig:Runtime_perf} , we also evaluate throughput in static settings, as shown in figure \ref{fig:throughput}. It's worth noting that these throughput metrics were obtained under continuous system tuning requests, which introduces unique considerations that may account for the observed discrepancies with the runtime speed-up results. We also documented the effects and insights related to tuning the structural parameters of the learned index, which are detailed in section~\ref{sec:eff}.

% \footnote{It is worth noting that after 30 steps, performance plateaus show diminishing returns from further tuning. This cutoff is set as extending beyond 30 steps, which leads to impractically long tuning times for most online users.}

%% Defending SOSD Data Representative Nature

\subsection{Baseline Methodologies}
\label{sec:baselines}
{\color{black}
Our evaluation sought to underscore the efficacy of \textsc{LITune} by contrasting it with various established tuning methodologies. Initially, the \textit{Default} method, utilizing unaltered system parameter settings from index designers or experts, set a fundamental baseline for comparative analysis. This is the adaptive configuration that responds to dynamic workloads and is designed by the authors of the corresponding indexes. \textit{Random Search} scanned the parameter space indiscriminately, while \textit{Grid Search} exhaustively tested predefined parameter combinations. Notably, \textit{Grid Search} does \emph{not} rely on expert defaults but instead uses a fixed parameter grid determined at the outset. \textit{Heuristic Search}, implemented via a simulated annealing kernel from OpenTuner~\cite{ansel2014opentuner}, pursued more focused exploration by leveraging existing expert heuristics. 
\textit{Sequential Model-Based Optimization (SMBO)} employed the TPE method~\cite{ozaki2020multiobjective,bergstra2013hyperopt}, effectively managing complex parameter spaces (an approach similar to OtterTune~\cite{van2017automatic}). 
Lastly, we incorporated a \textit{vanilla DDPG-based tuner}~\cite{lillicrap2015continuous}, referred to as \textbf{"DDPG"}, pretrained and fine-tuned with the same data as \textsc{LITune}, to demonstrate that direct RL pipelines from other domains (e.g., DBMS~\cite{zhang2019end}) may offer moderate gains when embedded in our framework, but are less effective overall than \textsc{LITune}. 
}

\begin{figure}[htbp]
\centering
    \includegraphics[width=0.8\linewidth]{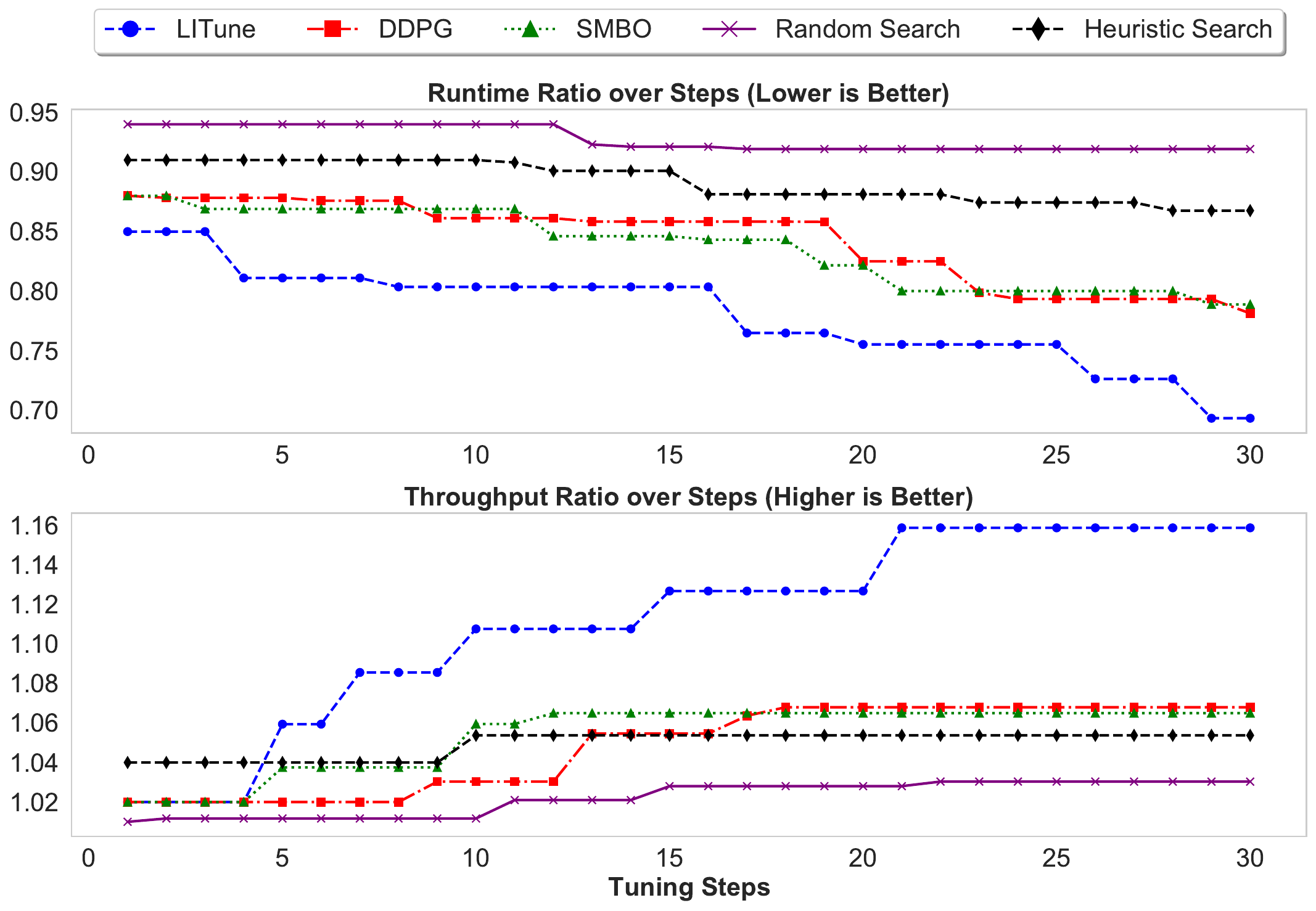}
\vspace{-8pt}
\caption{Tuning efficiency--Performance as tuning steps increase. \textbf{Above:} runtime ratio (best found vs. default settings). \textbf{Below:} throughput ratio (best found vs. default settings).}
\vspace{-8pt}
\label{fig:tuning}
\end{figure}

\vspace{-5pt}
\subsection{Results Overview}
\label{sec:results_overview}

Our comprehensive experimentation yielded results strongly in favor of \textsc{LITune}, with the method outperforming all baseline methods under varying conditions. Notably, \textsc{LITune}'s advantage becomes especially pronounced whether the available tuning steps are restricted or extended, emphasizing its efficacy and efficiency in optimizing learned index types.
\begin{figure}[htbp]
\centering
    \includegraphics[width=1.0\linewidth]{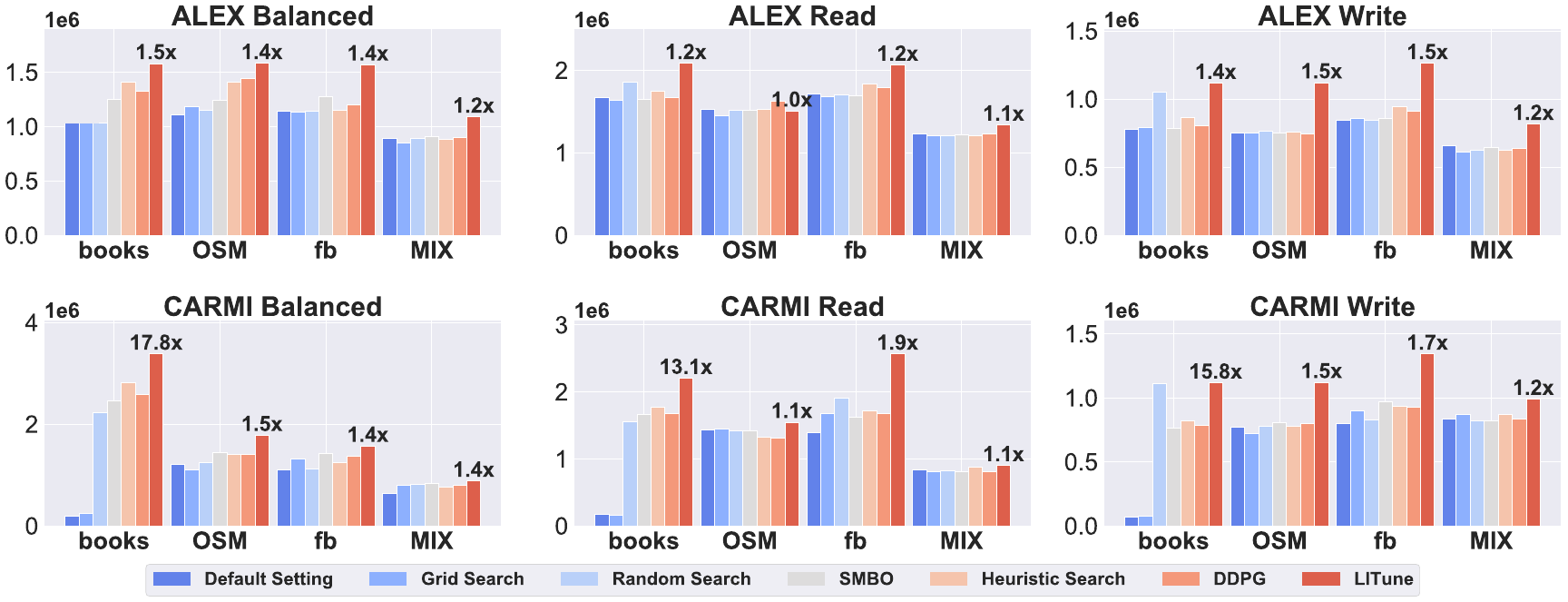}
\caption{Throughput performance (ops/sec) with extensive tuning. Performance improvements relative to default settings are marked.}
\vspace{-10pt}
\label{fig:throughput}
\end{figure}

\subsubsection{Tuning results and takeaways}
\label{sec:eff}
The values marked in Figure~\ref{fig:Runtime_perf} \& ~\ref{fig:throughput} show the performance gains compared against the default parameters as evidence of the effectiveness and need for a tuning system. We dive into some insight from the parameters of ALEX. Across all workloads, the thresholds for out-of-domain insertions exhibit significant increases compared to default parameters. Specifically, the minimum threshold shows an 80--100$\times$ increase and the maximum threshold ranges from 3--5$\times$, which suggests that ALEX can achieve better by "buffering" out-of-domain keys in each node prior to expansion. {\color{black}
Moreover, \textsc{LITune} adjusts workload-specific parameters such as the expected insertion fraction (otherwise fixed to a write-only workload) and toggles between approximate or exact model/cost computations. Specifically, ALEX benefits more from exact computations under balanced workloads, whereas read-heavy or write-heavy scenarios may favor approximate approaches. Additionally, \textsc{LITune} minimizes expansions and retrains, reducing retrains for the \textit{OSM} dataset under a balanced workload from 7196 to nearly zero. This outcome aligns with recent findings~\cite{sun2023learned,kim2024accelerating} emphasizing the overhead of retraining in learned indexes.
}

% Furthermore, ALEX benefits from having a larger maximum node size (64 MB compared to the default 16 MB).

% original texts
% In terms of workload-specific parameters, \textsc{LITune} is able to adjust the expected insertion fraction according to the workload, which by default is fixed to a write-only workload. Another interesting parameter is the trigger for approximate model and cost computation, which are both set to true by default. \textsc{LITune} shows for different workloads, ALEX benefits in the long run from computing the exact model and cost (which are more costly compared to approximations). Specifically, more balanced workloads benefit the most from exact model and cost computations. In other words, read-heavy or write-heavy workloads may benefit more from approximate model and cost computation. Lastly, we want to mention the states of ALEX, where \textsc{LITune} tries to minimize the number of expands and retrain states as much as possible. For the \textit{OSM} dataset under a balanced workload, it reduces the number of retrains from 7196 to close-to-0, which shows the cost of retrains. This insight coincides with findings from recently updated learned index research aiming to reduce retraining, the most costly operation in learned indexes.
% \vspace{-5pt}

% It is important to challenge the simplistic assumption that default parameter values are always sufficient.

\subsubsection{Radar Chart: Overall Evaluations among Tuning Methods}
\label{sec:eval_Radar}
% \vspace{-10pt}
\begin{figure}[htbp]
\centering
    \includegraphics[width = 0.86\linewidth]{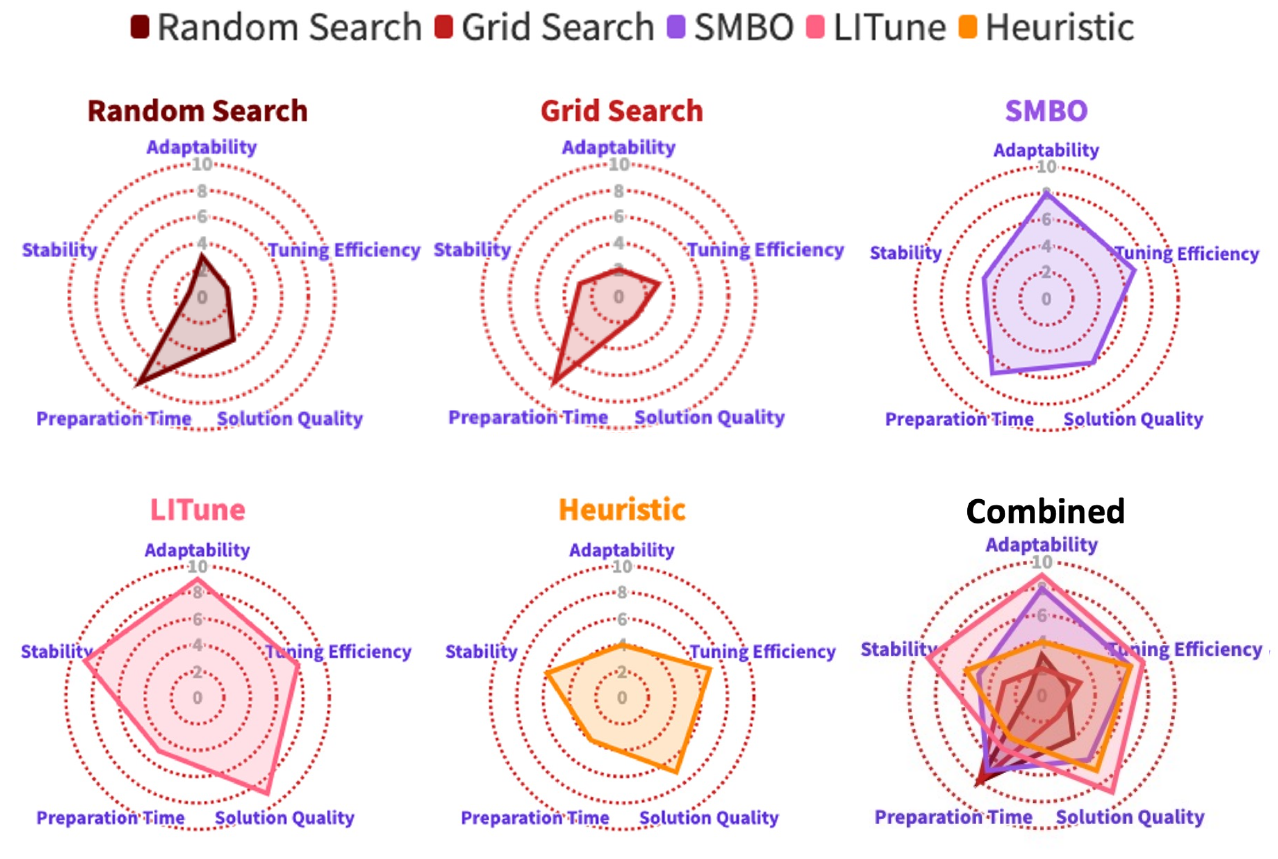}
\caption{Navigating the Parameter Tuning Landscape: \textsc{LITune} vs Other Tuning Methods}
\label{fig:Position}
\vspace{-10pt}
 % Adjust space below Figure
\end{figure}

The showcase in Figure~\ref{fig:Position}, based on 200 tuning trials using the MIX data distribution (the most complex) and a balanced workload on CARMI, quantitatively demonstrates the advantages of \textsc{LITune}. It provides a quick comparison of various tuning approaches across five attributes: Adaptability, Solution Quality, Stability, Tuning Efficiency, and Preparation Time. Scores are normalized on a scale from 0 to 9 to enable direct comparison. The results are illustrated in a radar chart, emphasizing the strengths and weaknesses of each method. Metrics are defined as: \textbf{Adaptability:} Lower runtime variances across scenarios indicates better adaptability.
\textbf{Solution Quality:} Higher average runtime performance signifies superior results.
\textbf{Stability:} More successful trials indicate higher stability.
\textbf{Tuning Efficiency:} Higher performance-to-tuning budget ratios show better efficiency.
\textbf{Preparation Time:} Shorter setup and pre-to-tuning time are preferred.

\subsubsection{Comparative E2E Performance of \textsc{LITune}} 
\label{sec:E2E}

{\color{black}
\textsc{LITune} demonstrates strong End-to-End (E2E) performance across diverse datasets and workloads, leveraging Meta-RL, the ET-MDP solver, and O2 System components. As shown in Figure~\ref{fig:tuning}, on the MIX dataset with a Balanced workload on ALEX, it rapidly achieves over 60\% of the optimal result within 10 steps, while others require at least twice as many. Its tuning efficiency---query runtime improvements per step---is about $2\times$ that of SMBO and DDPG, and $3\times$ that of random search, surpassing baselines from the very first recorded step. Only inference is required online, consuming just seconds per step on modern GPUs. Figure~\ref{fig:Runtime_perf} highlights CARMI’s notable optimization headroom, with over 90\% runtime reduction (compared to 30--40\% for ALEX). Complex datasets like MIX and OSM pose additional challenges, yet \textsc{LITune} consistently excels. 
}

When looking into tuning methods, \textit{Grid Search} and \textit{Random Search} display contrasting outcomes. \textit{Grid Search} consistently under-performs across all tuning budget scenarios, illustrating the challenges of exhaustively navigating extensive parameter spaces, and thus is not included in Figure~\ref{fig:tuning}. In contrast, \textit{Random Search} occasionally reaches optimal or near-optimal configurations under more generous budgets despite its inherent performance variability. \textit{SMBO}, while generally effective, can under-perform due to its reliance on historical evaluations, which may lead it into sub-optimal search areas influenced by noise or model discrepancies. Moreover, \textit{SMBO} risks venturing into system risk areas, potentially wasting the tuning budget without efficient exploration. \textit{Heuristic Search} maintains stable and reasonable performance but requires specific heuristic designs, which may not be universally applicable across all systems or scenarios. The performance of \textit{Heuristic Search} is notable in the ALEX scenario due to effective heuristic discovery. Yet, it faces limitations in the CARMI scenario where appropriate insights into structures and parameter impacts are lacking.

\textit{Failure of DDPG-Tuner:} As shown in Figure~\ref{fig:Runtime_perf} and Figure~\ref{fig:throughput}, vanilla DDPG only matches the performance levels of SMBO and heuristic searches, lagging 10-15\% behind \textsc{LITune}. The under-performance of a vanilla RL tuner using DDPG, compared to \textsc{LITune}, can be attributed to several key factors. Firstly, DDPG-Tuner often fails to capture the complex dependencies and dynamics within the tuning environment due to its simpler reinforcement learning model. This leads to underfitting, where the tuner is unable to generalize well from its training data to new or unseen scenarios. In contrast, \textsc{LITune} integrates advanced Meta-RL and the O2 system, which not only accelerates learning adaptations but also enhances its predictive accuracy by utilizing historical tuning experiences. Furthermore, DDPG-Tuner lacks the advanced memory and learning mechanisms of \textsc{LITune}, making it less effective in handling diverse tuning tasks, leading to sub-optimal decisions and reduced performance in complex environments.

\subsubsection{Tuning and Training Costs of \textsc{LITune}}
\label{sec:tuning_training}

\begin{table}[t]
\captionsetup{font=small,labelfont={bf,color=black}}
\caption{Training and tuning cost comparison across different methods. LITune-X denotes X\% sampling ratio, and Col 3-Col 6 indicate tuning time required to achieve target performance improvements.}
\vspace{-8pt}
\footnotesize
\label{tab:tuning_costs}
\centering
\begin{tabular}{lccccccc}
\toprule
\textbf{Method} & \textbf{Training} & \multicolumn{4}{c}{\textbf{Tuning Time}} & \multicolumn{1}{c}{\textbf{Best Perf.}} \\
\cmidrule(lr){3-6}
& & \textbf{-5\%} & \textbf{-10\%} & \textbf{-20\%} & \textbf{-45\%} & \multicolumn{1}{c}{(default 403s)} \\
\midrule
Grid Search & - & 32m & 1.0h & - & - & → 360s \\
Heuristic Search & - & 15s & 35s & 5m & - & → 312s \\
SMBO & - & 12s & 25s & 3m & - & → 314s \\
DDPG (\cite{lillicrap2015continuous,mo2023learning}) & 12h & 28s & 35s & 45s & - & → 326s \\
\midrule
LITune-0.1\% & 5h & 6s & 12s & 18s & 25s & → 288s \\
LITune-1\%(ours) & 6h & 8s & 15s & 22s & 28s & → 212s \\
LITune-10\% & 7h & 12s & 20s & 26s & 32s & → 211s \\
LITune-Full & 12h & 18s & 25s & 32s & 38s & → 208s \\
\bottomrule
\end{tabular}
\vspace{-10pt}
\end{table}

\checkcomment{
While \textsc{LITune} offers significant performance improvements, managing training and tuning costs is essential for practical deployment. Under a scaling workload with 400M balanced queries on ALEX using OSM data, Table~\ref{tab:tuning_costs} presents a comprehensive comparison of tuning overhead across different methods. We evaluate \textsc{LITune} under different sampling rates to demonstrate the effectiveness of our sampling strategy. For instance, to achieve a 20\% runtime reduction, \textsc{LITune} with 1\% sampling requires only 22 seconds of tuning time, compared to 25 seconds for vanilla DDPG and several minutes to hours for traditional approaches. We choose the 1\% sampling rate as it achieves nearly identical performance (212s vs 208s) to LITune-Full while having training and tuning overhead close to LITune-0.1\%, which sacrifices too much performance (288s vs 212s). Our results also showcase the highest attainable performance for various tuning methods when given substantial tuning budgets (1000s), while noting \textit{Grid Search}'s limitation of becoming computationally infeasible due to its expansive search domain.

The training phase represents a one-time investment in computational resources, where \textsc{LITune} develops generic tuning strategies across various index scenarios. As shown in Table~\ref{tab:tuning_costs}, \textsc{LITune}'s training time with 1\% sampling (6 hours) is half that of DDPG used in RusKey~\cite{mo2023learning} (12 hours), while providing superior tuning capabilities - achieving 47\% runtime reduction compared to DDPG's 19\%. Furthermore, our O2 System (detailed in Section~\ref{sec:O2}) allows instant model updates for real-time applications without additional training overhead. 
}

\vspace{-5pt}
\subsubsection{Adaptability of \textsc{LITune}}
\label{sec:adapt}

\begin{figure}[htbp]
\centering
    \includegraphics[width=0.9\linewidth]{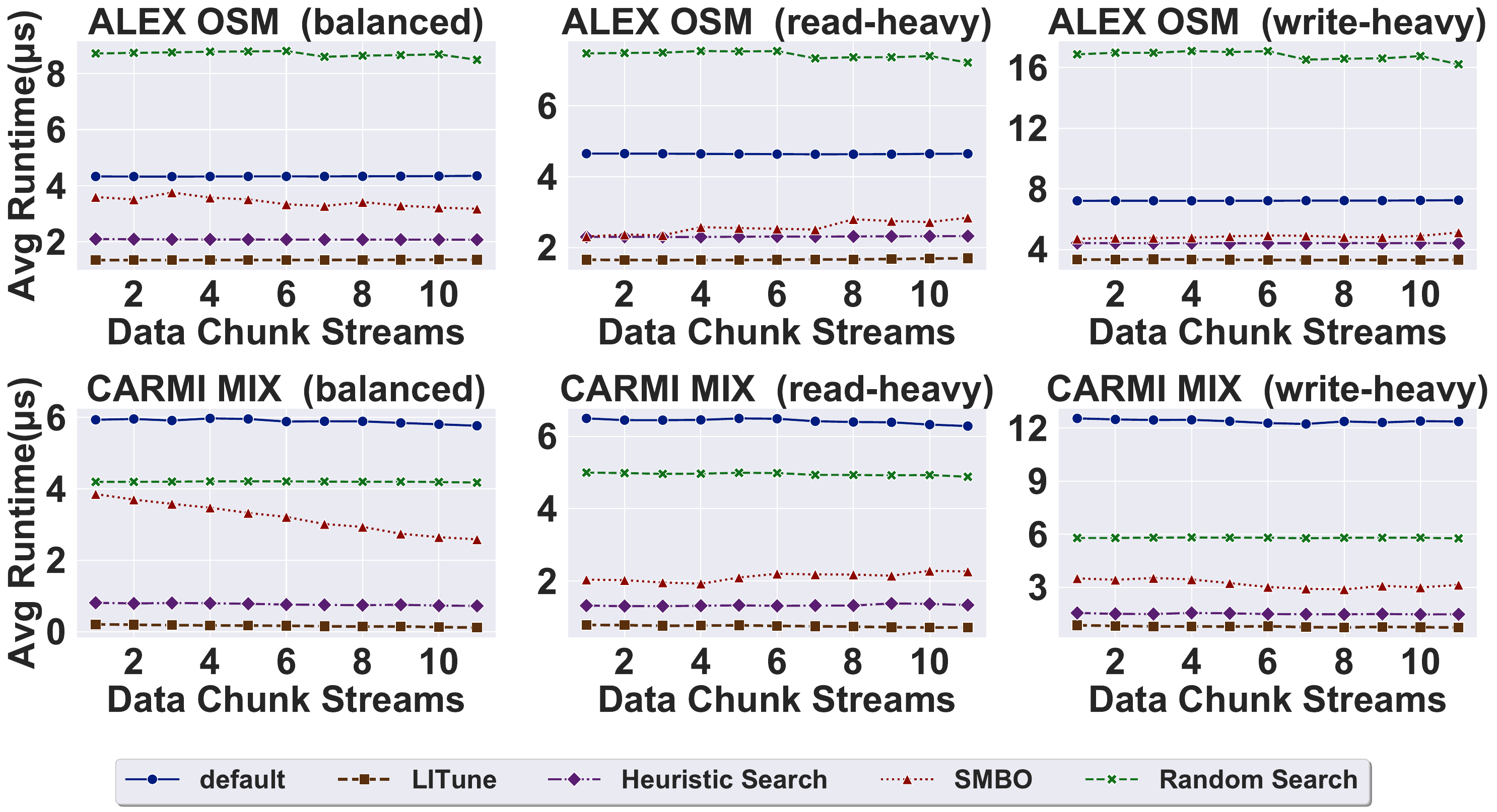}
\caption{Online and continuous tuning performance (averaged over operation tuples) in data streams (ALEX+OSM and CARMI+MIX)}
\vspace{-5pt}
\label{fig:DS}
\end{figure}

\textsc{LITune} consistently exhibits superior performance and adaptability across varied queries, data distributions, and data shifts, as demonstrated in Figures~\ref{fig:Runtime_perf}, \ref{fig:throughput}, and \ref{fig:DS}.

This robust performance amid diverse and shifting scenarios can be ascribed to the foundational Deep Reinforcement Learning (DRL) framework, which enables \textsc{LITune} to continually refine its policies and swiftly adapt its parameter settings, ensuring optimal performance amidst varied query contexts and data scenarios. The \textsc{LITune} not only promotes intelligent and dynamic exploration and exploitation of the parameter space but also facilitates quick convergence to optimal or near-optimal configurations, making it especially adept at navigating complex, heterogeneous, and dynamically evolving data and query environments, thereby addressing \textbf{\fcolorbox{blue}{white}{\hyperlink{c2}{C.2}}}. Specifically, \textsc{LITune} demonstrates:

\textit{1. Consistency across Query Types:} Different rows in Figure~\ref{fig:Runtime_perf} correspond to diverse query types (B, RH, WH) when read vertically. Notably, \textsc{LITune}’s framework efficiently navigate through different query types, adapting its parameter settings in real time to ensure optimal performance amidst shifting query contexts.

\textit{2. Versatility across Data Distributions:} Different columns in Figure~\ref{fig:Runtime_perf} correspond to varied data distributions (OSM, books, fb, MIX) when read horizontally. \textsc{LITune} also showcases adaptability to varied data distributions, adeptly managing complex scenarios like the MIX distribution by autonomously identifying and applying optimal parameter configurations. 

\textit{3. Robustness amidst Data Shifts:} As evident in Figure~\ref{fig:DS}, \textsc{LITune} sustains high performance across continuous data chunks in online tuning, quickly adapting to evolving data distributions without necessitating re-initialization. It is important to note that we did not include DDPG or vanilla-DRL methods for comparison here, as their extreme instability in handling dynamic scenarios without tailored designs made them unsuitable for this evaluation.

\subsection{Ablation Study}

\subsubsection{Effects of O2 System}

\begin{figure}[htbp]
\centering
    \includegraphics[width=0.88\linewidth]{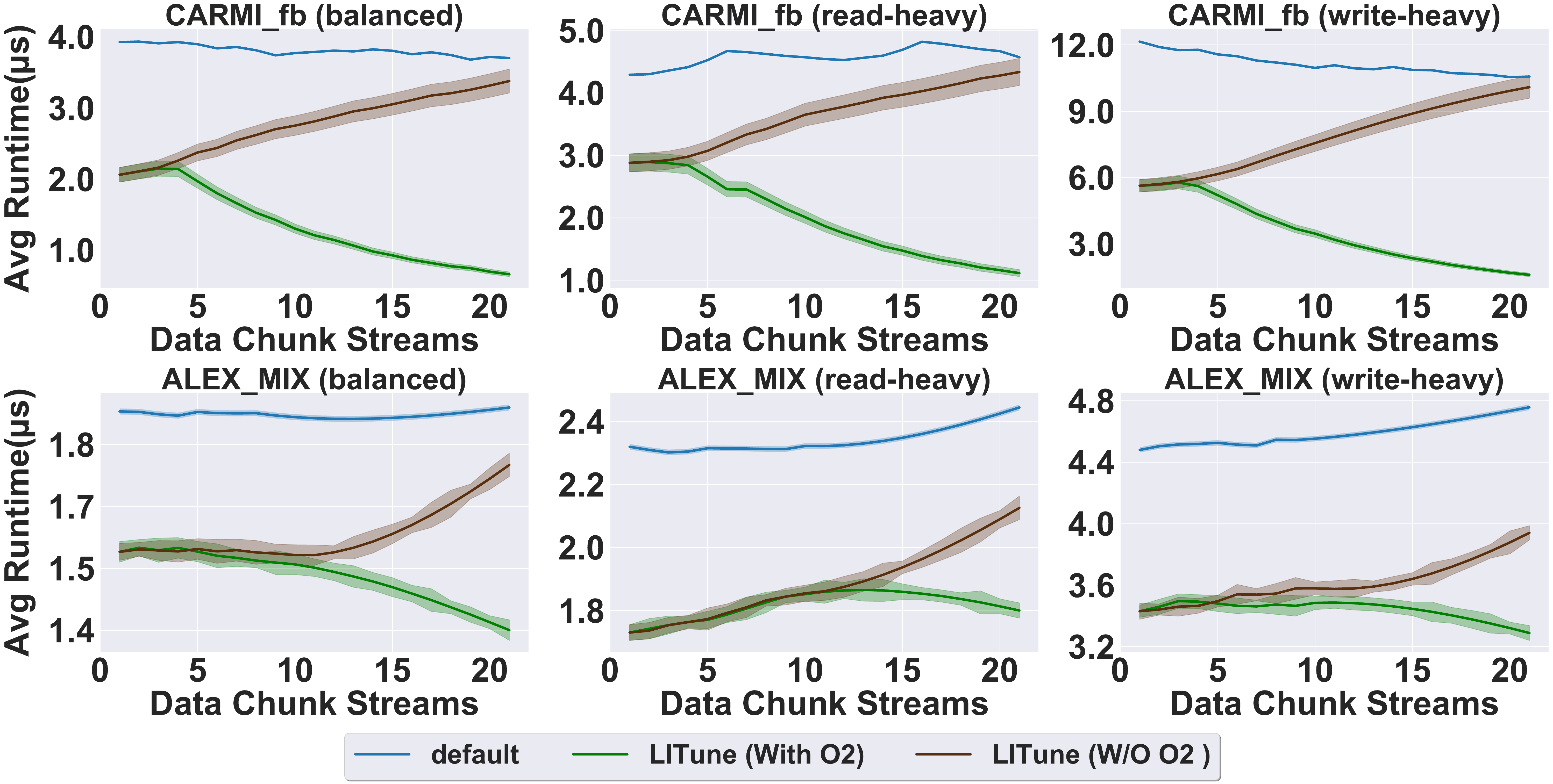}
\caption{Benefits of the O2 system in online and continuous tuning, comparing with pre-trained models W/O O2 system (CARMI+fb and ALEX+MIX), averaged over operation tuples.}
\label{fig:O2}
\vspace{-10pt} % Adjust space below Figure
\end{figure}

\begin{figure}[htbp]
\centering
\includegraphics[width=0.88 \linewidth]{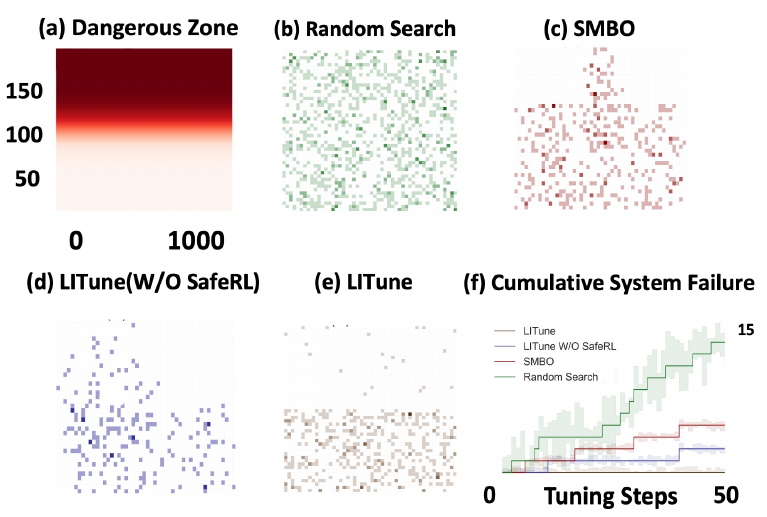}
\caption{Exploring Parameter Spaces Across Tuning Methods(ALEX + OSM + Balanced).}
\label{fig:risk_mitigation}
% \vspace{-10pt}
\end{figure}

% \vspace{-10pt}

\textbf{Resilience to Unseen Data:} Figure \ref{fig:O2} shows that the O2 system's online component quickly adapts to new trends using a sliding window of recent queries, enhancing resilience to unforeseen data and ensuring adaptability.

\textbf{Handling Data Distribution Shifts:} Demonstrated in the MIX dataset with ALEX (See Figure \ref{fig:O2}), the O2 system adeptly handles data distribution shifts. The offline tuner, enriched with diverse training data, robustly supports varied distributions. When the online model detects significant shifts, a threshold in the divergence measure initiates a model swap, ensuring continuous adaptation.

\textbf{Consistency in Performance:} As demonstrated in Figure \ref{fig:O2}, \textsc{LITune} equipped with the O2 system consistently outperforms versions without it. This superior performance is attributed to its hybrid approach: the online model rapidly adapts to new conditions, while the offline model, trained on a wider array of data scenarios, seamlessly maintains performance. This synergy ensures dependable and stable operation across diverse data environments.

% The use of an initial soft rollout further fortifies the system against unexpected performance drops. By allowing the offline model to handle a fraction of the queries initially, the system can assess the offline model's capabilities before fully implementing it. This strategy offers a safety net, ensuring that the new model doesn't introduce more errors than it aims to correct.

\subsubsection{Safe Tuning}
\label{sec:safe_tune}
As mentioned in the Introduction, exploring parameter spaces can enhance system performance but also introduces risks, especially when parameters significantly impact system reliability and stability during tuning and operation (\textbf{\fcolorbox{blue}{white}{\hyperlink{c3}{C.3}}}).

\begin{figure}[htbp]
\vspace{-5pt}
\captionsetup{labelfont={bf,color=black}}
\centering
\includegraphics[width=0.95\linewidth]{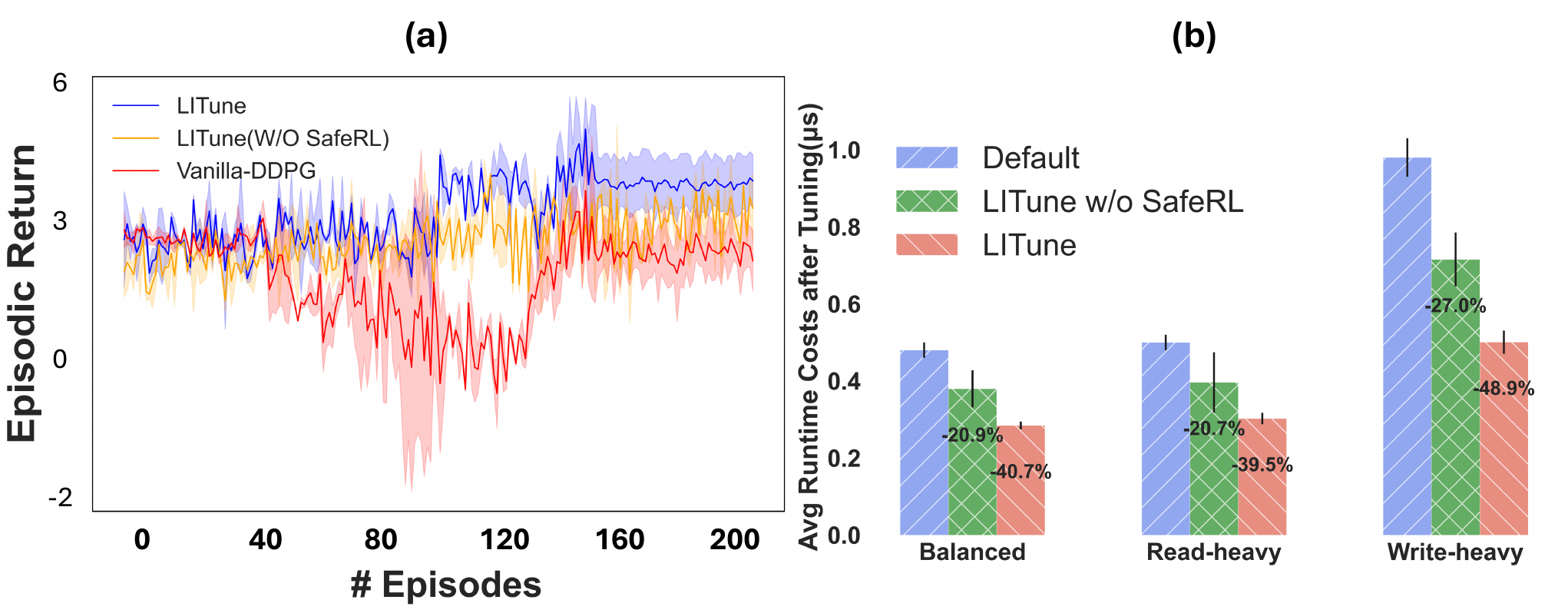}
\caption{(a) Training stability comparison between LITune variants with and without Safe-RL (ALEX). (b) End-to-end runtime performance of LITune variants after tuning on the MIX dataset for ALEX.}
\label{fig:ab_curves}
\vspace{-5pt}
\end{figure}

\checkcomment{Figure~\ref{fig:ab_curves}(a) demonstrates the importance of our Safe-RL module through training performance comparison. Without Safe-RL, the reward signals exhibit high volatility during the latter stages of training due to severe penalty signals triggered by system terminations (e.g., infinite loops, memory crashes) from aggressive parameter exploration. In contrast, \textsc{LITune} with Safe-RL maintains stable reward improvement throughout training, achieving both better final performance and more consistent learning progress compared to vanilla DDPG. Besides, within 200 epochs, we found that the vanilla DDPG cannot learn a converged policy and requires further training. This stability difference is particularly pronounced after the 100th training episode, where \textsc{LITune} without Safe-RL shows large reward fluctuations from frequent system failures, while the Safe-RL version maintains steady improvement by proactively avoiding parameter combinations that could lead to system termination. As shown in Figure~\ref{fig:ab_curves}(b), this training stability translates directly to better end-to-end performance from trained policies: In average, \textsc{LITune} with Safe-RL achieves 30\% lower runtime with 60\% less variance compared to the version without Safe-RL, demonstrating that safer exploration leads to both better and more reliable tuning outcomes}

Figure~\ref{fig:risk_mitigation} further illustrates the benefits of our safety-aware design. Figures~\ref{fig:risk_mitigation}(a)--(e) show how four methods explore two critical parameters: \textit{kMaxOutOfDomainKeys} and \textit{kOutOfDomainToleranceFactor}, which are crucial under specific configurations (\textit{fanoutSelectionMethod = 1}, \textit{splittingPolicyMethod = 1}, \textit{allowSplittingUpwards = True}). The red-highlighted \textbf{Dangerous Zone} in subfigure (a) marks parameter regions prone to causing system instabilities and degraded performance, emphasizing the need for a tuner like \textsc{LITune}. subfigure (f) presents the cumulative number of index system failures (e.g., infinite loops, memory issues) during tuning over five trials, highlighting the effectiveness of our safety design.

\textit{Random Search}, lacking predictive or memory capabilities, explores the parameter space indiscriminately, entering risky zones and introducing instability into the learned index structures. \textit{SMBO} incorporates a model to partially understand the parameter space, offering semi-protected exploration but still occasionally ventures into risky regions due to limited learning capacity. It tends to focus sampling in specific areas, sometimes within dangerous zones, increasing the risk of system instability. \textit{\textsc{LITune} (without Safe-RL)}, lacking a context-aware design, also struggles with risky areas but performs better than other baselines because its reward function penalizes long runtimes, raising awareness of safe tuning regions.

In contrast, \textsc{LITune} embeds stability into exploration by leveraging historical knowledge to avoid the \textbf{Dangerous Zone}, ensuring reliable and stable index structures.

% by steering clear of risky parameter configurations.

% In contrast, \textit{LITune} actively embeds stability into the exploration process, leveraging historical knowledge to navigate and avoid previously identified dangerous areas. By carefully steering clear of the \textbf{Dangerous Zone}, \textsc{LITune} ensures that learned index structures remain reliable and stable by strategically avoiding parameter configurations known to produce undesirable outcomes.

\vspace{-5pt}
\section{Discussion and conclusion}
\label{sec:con}

\checkcomment{
\textsc{LITune} is effective across diverse learned indexes, though its impact varies with parameter space size and interdependencies. By treating tuning as a black-box problem, it first identifies optimal configurations, then maps them back for analysis. Our studies indicate RL tolerates certain redundant parameters without harming performance, provided dimensionality stays manageable.  While we currently rely on domain expertise to pinpoint key parameters, a systematic approach could automate this. Techniques like checkpointing and caching successful configurations further reduce retraining overhead as workloads evolve. In future work, preliminary sensitivity analysis can prune superfluous parameters for new or evolving indexes.
}

Furthermore, while \textsc{LITune} shows considerable promise in learned index optimization, several areas warrant further exploration. Like other RL-based methods, it occasionally faces convergence and stability issues during training, highlighting the need for future research to address these challenges to improve system reliability. Additionally, accelerating offline training (e.g., using parallelization) could further enhance the efficiency of \textsc{LITune}'s deployment. Despite these challenges, \textsc{LITune}'s innovative approach, combining Meta-RL and context-aware strategies, sets a new standard by improving system efficiency and query performance while ensuring stability through proactive risk management.

% \bibliographystyle{ACM-Reference-Format}

\bibliographystyle{ACM-Reference-Format}
\bibliography{sample-base}

\end{document}